\documentclass[preprint,12pt]{elsarticle}

\usepackage{amssymb}
\usepackage{amsthm}

\usepackage{subcaption}
\usepackage{amsmath,mathrsfs}
\usepackage{mathtools}
\usepackage{amssymb}
\usepackage{float}
\usepackage{tikz}
\usepackage{tensor}
\usepackage{makecell}
\usetikzlibrary{tikzmark}
\usepackage{tabstackengine}
\usetikzlibrary{decorations.pathreplacing}
\usepackage[title]{appendix}
\usepackage{lscape}

\journal{Nuclear Physics B}

\begin{document}

\begin{frontmatter}



\title{Effective Lagrangian for Non-Abelian Two-Dimensional Topological Field Theory}


\author[inst1,inst2]{Pongwit Srisangyingcharoen}
\ead{pongwit.srisangyingcharoen@durham.ac.uk, pongwits@nu.ac.th}
\affiliation[inst1]{organization={Centre for Particle Theory, University of Durham},
            city={Durham},
            postcode={DH1 3LE}, 
            country={United Kingdom}}

\author[inst1]{Paul Mansfield}
\ead{p.r.w.mansfield@durham.ac.uk}
\affiliation[inst2]{organization={The Institute for Fundamental Study, Naresuan University},
            city={Phitsanulok},
            postcode={65000}, 
            country={Thailand}}

\begin{abstract}
We develop a systematic approach to obtain an effective Lagrangian for $2D$ non-Abelian topological BF theory. A general expression is presented in a diagrammatic representation  containing solely scalar fields. Expressions for the $SU(2)$ and $SU(3)$ effective actions are explicitly stated. In the case of $SU(2)$, we show that the effective action can be interpreted as a winding number. By using the $SU(2)$ effective action, the partition function on a sphere for $SU(2)$ Yang-Mills theory is calculated. Moreover, we generalise the theory to include a source term for the gauge field as well as calculate the vacuum expectation value of the Wilson loop based on the effective theory.
\end{abstract}


\end{frontmatter}

\tableofcontents

\section{Introduction}
Over the past few decades, the study of topological field theories has been important for both mathematics and physics. The key feature of the theories is that observables depend only on the global structure of the space where the theories are defined. Among the field theory models of topological type, the BF theory is of particular interest in many area in physics. It plays an important role as an alternative theory of gravity \cite{CHAMSEDDINE1990595,Freidel:2012np,Guo:2002yc,Smolin:1995vq,WITTEN198846,Freidel:1999rr,Celada:2016jdt} and quite recently, it has gained much attention for its topological effects in condensed matter physics \cite{Vishwanath:2012tq,Marzuoli_2012,Palumbo:2013gxa,Cho:2010rk,Thakurathi:2020veg,Blasi:2011pf,You2019FractonicCA}. 

In this paper, we will obtain the effective Lagrangian purely in terms of  the scalar field $\phi$ for the BF theory resulting in a theory that is both gauge and Weyl invariant. Our motivation for constructing this model is to be able to generalise the results of \cite{Mansfield:2011eq}, \cite{Edwards:2014cga}, and \cite{Edwards:2014xfa} in which it was shown that the Wilson loop for $D$-dimensional Abelian gauge theories could be obtained from a spinning string theory in which the physical degrees of freedom described by the string are electric lines of force. The interaction in this model is not the usual splitting/joining interaction of fundamental string theory but rather (the supersymmetrisation of) a contact interaction that is supported on self-intersections of the string. If the string target space co-ordinates are denoted by $X^\mu(\xi^1,\xi^2)$ where $\xi^i$ are world-sheet parameters for an open string with fixed boundary curve $C$ then $d\Sigma^{\mu\nu}(\xi)=d^2\xi'\,\epsilon^{ij}
\partial_i X^\mu(\xi) \,
\partial_j X^\mu(\xi)$ is an element of area in target space and the contact interaction takes the form
\begin{equation}
\int d\Sigma^{\mu\nu}(\xi)\,\,
\delta^D\left(X(\xi)-X(\xi')\right)\,
\,d\Sigma_{\mu\nu}(\xi')
\end{equation}
with a coefficient proportional to the square of the electric charge. Averaging this over fluctuating world-sheets results in the gauge-field propagator connecting points $X(\xi)$ and $X(\xi')$ when they are on the boundary $C$. We would like to generalise this to non-Abelian gauge theory so at the very least we need to introduce Lie algebra-valued world-sheet degrees of freedom $\phi(\xi)$ to try to reproduce the Lie algebra structure of Yang-Mills propagators
\begin{equation}
\int d\Sigma^{\mu\nu}(\xi)\,\,
\delta^D\left(X(\xi)-X(\xi')\right)\,
\,d\Sigma_{\mu\nu}(\xi')\,{\rm tr}\left( \phi(\xi)\,\phi(\xi')\right)\,.
\end{equation}
As a consequence of the $\delta$-function this interaction is invariant under gauge transformations which are functions on target-space if $\phi$ transforms as $\phi(\xi)\rightarrow g\left(X(\xi)\right)\,\phi(\xi)\,g^{-1}\left(X(\xi)\right)$. To construct a string theory describing non-Abelian lines of force we need a Lagrangian to describe the dynamics of $\phi$. This has to be gauge-invariant to preserve the space-time gauge invariance of the contact interaction and Weyl invariant to satisfy the usual organising principle of string theories. It also has to generate the extra interactions of non-Abelian gauge theories which are absent from Abelian ones. A candidate for the dynamics of $\phi$ is the effective Lagrangian for the BF theory we will calculate in this paper. We shall not address here the issue of whether the resulting theory does indeed generate the self-interactions of Yang-Mills theory but simply concentrate on deriving the model and addressing the examples of gauge groups $SU(2)$ and $SU(3)$ relevant to the Standard Model.

A BF theory is a diffeomorphism-invariant gauge theory. On a $D$-dimensional manifold $\mathcal{M}$ $(D\geq 2)$ with structure group, a Lie group $G$, the classical action of the non-Abelian BF theory takes the form
\begin{equation}
    S=2\int_\mathcal{M} \text{tr}(B \wedge F) \label{BF}
\end{equation}
where $B$ is a $(D-2)$-form in the fundamental representation of $G$. $F$ is a curvature 2-form of a connection 1-form $A$ defined by $F=dA+g[A,A]$. The trace implies a scalar product in the algebra. Notice that the action is topologically invariant because it is independent of the metric. The equations of motion with respect to $B$ and $A$ are
\begin{equation}
    F=0 \qquad \text{and} \qquad d_A B=0
\end{equation}
where $d_A$ is a covariant derivative defined as $d_A=d+g[A,\ \ ]$. The action is invariant under local gauge transformation with gauge parameter $\omega$ as
\begin{equation}
    \delta A=d_A \omega \qquad \text{and} \qquad \delta B=[B,\omega]+d_A \eta. \label{gauge trans}
\end{equation}
The field $\eta$ is a $(D-1)$-form corresponding to the non-Abelian symmetry of the $B$ field, namely $B$ symmetry which only appears when $D \geq 3$.

In the case $D=2$, one can express (\ref{BF}) as
\begin{equation}
    S=2\int_\mathcal{M} d^2\xi \ \epsilon^{ij} \text{tr}(\phi \mathcal{F}_{ij}) \label{Topological field action}
\end{equation}
or equivalently,
\begin{equation}
    S[\phi,\mathcal{A}]=\int_\mathcal{M} d^2\xi \ \Bigg(ig\mathcal{A}\indices{_i_A}\mathcal{A}\indices{_j_B}  f\indices{^A^B^C}\phi\indices{_C}-2\partial_i\phi_A \mathcal{A}^A_j   \Bigg)\epsilon^{ij} \label{topological field action2}
\end{equation}
where $\mathcal{F}_{ij}=\partial_i\mathcal{A}_j-\partial_j\mathcal{A}_i+g[\mathcal{A}_i,\mathcal{A}_j]$ is the field-strength of the gauge field $\mathcal{A}$. Both fields $\phi$ and $\mathcal{A}$ are elements of a non-Abelian group and  so can be written in terms of a set of generators $\{T^R\}$ as $\phi=\phi_RT^R$ and $\mathcal{A}=\mathcal{A}_RT^R$.\footnote{tr($T^AT^B$)=$\frac{1}{2}\eta^{AB}$ and $[T^A,T^B]=if\indices{^A^B_C}T^C$} To obtain ({\ref{topological field action2}}), the boundary term, i.e. $\int d^2\xi \partial_i( \phi \cdot\mathcal{A}_j)\epsilon^{ij}$, is assumed to vanish. Notice that in two dimensions, the $B$ field is a 0-form, thus, it is natural  to replace it with the scalar field $\phi$.

The action (\ref{Topological field action}) has a close connection to the Yang-Mills action in two dimensions as they are equivalent in the zero coupling constant limit \cite{witten1991,Witten:1992xu}. This can be seen by adding a quadratic term with coupling constant $e$ to the action and then integrating out the field $\phi$ in the path integral below using Gaussian integration
\begin{align}
  &\int D\mathcal{A} D\phi \exp\bigg( 2\int_\mathcal{M} d^2\xi \ \bigg( \epsilon^{ij} \text{tr}(\phi \mathcal{F}_{ij}) +e^2 \sqrt{g} \text{tr}(\phi \phi) \bigg) \bigg) \nonumber \\
  &=\int D\mathcal{A} \exp\bigg( \frac{1}{2e^2}\int_\mathcal{M} d^2\xi \sqrt{g}\text{tr}(\mathcal{F}_{ij}\mathcal{F}^{ij}) \bigg). \label{YM}
\end{align}

The outline of this paper is as follows. In section two, an explicit calculation for the $SU(2)$ effective BF theory is shown. In section three we use the result to give a new derivation of the known expression for the partition function on a sphere of $SU(2)$ Yang-Mills theory. The calculation for obtaining the effective theory is generalised for arbitrary Lie algebras in section four. In section five, we construct a set of diagrams to represent the ingredients appearing in the effective Lagrangian in order to aid our calculation. Using the result found in section five we present the explicit form for the $SU(3)$ effective Lagrangian in section six. Finally, in section seven, we investigate the BF model with a source term for a gauge field $\mathcal{A}$ as well as calculate the expectation value of the Wilson loop in the effective theory.
\section{$SU(2)$ Effective BF Theory}
We begin our calculation with the simplest model for the non-Abelian two-dimensional topological field theory, i.e. the BF theory for $SU(2)$. The partition function for this theory is defined as
\begin{equation}
        Z=\frac{1}{\text{Vol}}\int D\phi D\mathcal{A} \ e^{-S[\phi,\mathcal{A}]} \label{path integral}
\end{equation}
where $S[\phi,\mathcal{A}]$ is expressed in (\ref{topological field action2}). The functional integral is divided by the volume of the gauge symmetry which is denoted by Vol.

To obtain an effective theory for the scalar field $\phi$, the gauge field $\mathcal{A}$ needs to be integrated out. For that purpose, we express all fields in terms of a set of orthonormal bases, i.e. $\hat\phi, \hat{E}_+$, and $\hat{E}_-$,  as
\begin{equation}
    \phi^A =\varphi \hat\phi^A \qquad \text{and} \qquad \mathcal{A}^A_i=\chi_i\hat\phi^A+a^+_i\hat{E}^A_+ +a^-_i\hat{E}^A_-. \label{expand}
\end{equation}
Note that these bases are $\xi$-dependent. They are defined throughout the manifold point by point. Obviously, we have chosen a unit vector  $\hat\phi$  to align in the direction of $\phi$ at each point. In terms of the usual cross products, the $\xi$-dependent bases give the following relations:
\begin{equation}
    \hat{\phi}\times\hat{E}_+=\hat{E}_+, \quad \hat{E}_+\times \hat{E}_-=\hat\phi, \quad \hat{E}_-\times \hat\phi=\hat{E}_-.\label{cross}
\end{equation}

Substituting (\ref{expand}) into  (\ref{topological field action2}), the action takes the form
\begin{equation}
    S[\phi,\mathcal{A}]=\int_\mathcal{M} d^2\xi \ \Bigg(2ig\varphi\  a^+_i a^-_j -2\partial_i\phi_A \chi_j \hat{\phi}^A -2\partial_i\phi_A a^+_j \hat{E}^A_+ -2\partial_i\phi_A a^-_j \hat{E}^A_- \Bigg)\epsilon^{ij}.
\end{equation}
To obtain the first term, the relations (\ref{cross}) were utilised. Note that the structure constant $f^{ABC}$ is equal to $\epsilon^{ABC}$ for $SU(2)$.

Rewriting all the fields using (\ref{expand}), the measure $D\mathcal{A}$ now turns into $D\chi Da^+Da^-$. Integrating out $\chi$ would generate a constraint via the Dirac delta function as 
\begin{align}
    \int D\chi_i \ \text{exp}\ \bigg(\int_\mathcal{M} d^2\xi \ 2\hat\phi^A\partial_i\phi\indices{_A} \chi_j \epsilon^{ij} \bigg)=\mathcal{N} \prod_{\forall \xi \in \mathcal{M}} \varphi^2 \delta^{(2)}(\underline{\partial}\varphi^2) =\mathcal{N} \prod_{\forall \xi \in \mathcal{M}}\delta^{(2)}(\underline{\partial}\varphi).\label{SU2 constraint}
\end{align}
This means $\varphi^2$ (equivalently $|\phi|^2$) is constant throughout the space $\mathcal{M}$.

To proceed with the path integration with respect to the field $a^\alpha_i$ with $\alpha=\pm$, it is better to change the spacetime coordinates $\xi^1$ and $\xi^2$ into complex coordinates which are defined by
\begin{equation}
    z=\xi^1+i\xi^2 \qquad \text{and} \qquad \bar{z}=\xi^1-i\xi^2. \label{complex coord}
\end{equation} 
In these new coordinates, the field $a^\alpha_i$ becomes  complex fields $b^\alpha$ where
\begin{equation}
    b^\alpha=\frac{1}{2}(a^\alpha_1-ia^\alpha_2) \qquad \text{and} \qquad \bar{b}^\alpha=\frac{1}{2}(a^\alpha_1+ia^\alpha_2). \label{complex b field}
\end{equation}

Therefore, the path integral (\ref{path integral}) takes the form
\begin{equation}
 Z=\frac{1}{\text{Vol}}\int D\phi Db D\bar{b} \prod_{\forall \xi \in \mathcal{M}}  \delta^{(2)}(\underline{\partial}\varphi) \ e^{-S[\phi,b,\bar{b}]} \label{integrate out b}
\end{equation}
where
\begin{equation}
    S[\phi,b,\bar{b}]=\int_\mathcal{M} d^2z \ \Bigg(-\bar{b}^\alpha2ig\varphi\epsilon_{\alpha\beta} b^\beta +2\bar\partial\phi_A \hat{E}^A_\alpha b^\alpha -2\partial\phi_A \hat{E}^A_\alpha \bar{b}^\alpha \Bigg).\label{action complex field}
\end{equation}
We can then use the Gaussian integration formula to integrate out the complex field $b$,
\begin{equation}
    \int Db D\bar{b} \ e^{-\int d^2z(-\bar{b}^\alpha M_{\alpha\beta} b^\beta+\bar{J}_\alpha b^\alpha+J_\alpha \bar{b}^\alpha)}=\mathcal{N}_0\frac{e^{-\int d^2z (\bar{J}_\alpha (M^{-1})^{\alpha\beta} J_\beta}}{\prod\limits_{\forall \xi}\text{det}(M)}. \label{gaussian}
\end{equation}
According to (\ref{action complex field}), it is not hard to see that
\begin{equation}
    M_{\alpha\beta}=2ig\varphi \epsilon_{\alpha\beta}, \quad J_\alpha=-2\partial\phi_A \hat{E}^A_\alpha, \quad \text{and} \quad \bar{J}_\alpha=2\bar\partial\phi_A \hat{E}^A_\alpha.
\end{equation}
Using the fact that $\epsilon^{ij}\epsilon_{ij}=2$, the inverse and the determinant of the matrix $M$ are
\begin{equation}
(M^{-1})^{\alpha\beta}=\frac{-i}{4g\varphi}\epsilon^{\alpha\beta}, \qquad \text{and} \qquad \det(M)=-4g^2\varphi^2.    \label{su2 det}
\end{equation}
Consequently, we can express the path integral as 
\begin{align}
 Z&\sim\int D\phi\prod_{\forall \xi \in \mathcal{M}} \frac{-i}{(g\varphi)^2} \delta^{(2)}(\underline{\partial}\varphi) \ \exp\bigg[-\int_\mathcal{M} d^2z \frac{i}{g\varphi} \bar{\partial}\phi\indices{_A}\partial\phi\indices{_B}(\hat{E}^A_\alpha\epsilon^{\alpha\beta}\hat{E}^B_\beta) \bigg].
\end{align}
We can rewrite the term $\hat{E}^A_\alpha\epsilon^{\alpha\beta}\hat{E}^B_\beta$ as
\begin{equation}
    \hat{E}^A_\alpha\epsilon^{\alpha\beta}\hat{E}^B_\beta=\hat{E}^A_+\hat{E}^B_--\hat{E}^B_+\hat{E}^A_-=(\hat{E}_+\times\hat{E}_-)_C\epsilon\indices{^A^B^C} \label{cross product}
\end{equation}
which can be evaluated using (\ref{cross}). As a result,  the cross product on the right-hand side is simply the unit vector $\hat\phi$. Thus, the effective action for two-dimensional SU(2) BF theory can be written as
\begin{equation}
    \int_\mathcal{M} d^2z \frac{i}{g|\phi|^2} \bar{\partial}\phi\indices{_A}\partial\phi\indices{_B}\phi_C\epsilon\indices{^A^B^C} \label{SU2 eff}
\end{equation}
or equivalently in the $(\xi^1,\xi^2)$ coordinates as
\begin{equation}
    \int_\mathcal{M} d^2\xi \frac{i}{2g|\phi|^2} \partial_i\phi\indices{_A}\partial_j\phi\indices{_B} \epsilon^{ij}\phi_C\epsilon\indices{^A^B^C}. \label{SU2 eff2}
\end{equation}

Now, let us give an interpretation of the effective action (\ref{SU2 eff2}). 
The effective action can be seen as a winding number (up to a constant). To see this, it needs to be noted that the unit vector $\hat\phi(\xi)$ maps a point on the manifold $\mathcal{M}$ into a point on $S^2$, i.e. $\hat\phi: \mathcal{M} \rightarrow S^2$. Furthermore, the integrand of the action (\ref{SU2 eff2}),
\begin{equation}
\frac{1}{2}\partial_i\hat\phi\indices{_A}\partial_j\hat\phi\indices{_B} \epsilon^{ij}\hat\phi_C\epsilon\indices{^A^B^C}, \label{area element}
\end{equation}
is the area element on the target space $S^2$. This can be seen as follows: the variations of the manifold coordinates $\delta\xi^1$ and $\delta\xi^2$ correspond to two infinitesimal tangent vectors $\delta\xi^1\partial_1\hat\phi$ and $\delta\xi^2\partial_2\hat\phi$ on $S^2$. The cross product of these two vectors has direction $\hat\phi$ and magnitude $\delta A$. Consequently, the triple product, $\delta\xi^1\delta\xi^2(\partial_1\hat\phi \times \partial_2\hat\phi)\cdot\hat\phi$, is basically an infinitesimal area on the target space $S^2$ as claimed.

The integration over all manifold coordinates $\xi$ of the integrand (\ref{area element}) yields the total area of the unit sphere times an integer corresponding to the winding number $n$ as 
\begin{equation}
   \frac{1}{2} \int_{S^2} d^2\xi  \partial_i\hat\phi\indices{_A}\partial_j\hat\phi\indices{_B} \epsilon^{ij}\hat\phi_C\epsilon\indices{^A^B^C} =4\pi n. \label{winding number}
\end{equation}
Note that the above term is proportional to the effective action (\ref{SU2 eff2}) as the magnitude of the field $\phi$, $|\phi|$, is constant due to the constraint (\ref{SU2 constraint}). 

\section{Partition Function for $SU(2)$ Yang-Mills Theory on Sphere}
It is well known that a general expression for partition function for $SU(N)$ Yang-Mills theory on a sphere is 
\begin{equation}
    Z_{\text{YM}}(A)=\sum_R (d_R)^2\exp\big( -e^2A C_2(R) \big)
\end{equation}
where $A$ is an area of the sphere and $R$ is an irreducible representation of $SU(N)$. $d_R$ and $C_2(R)$ are the dimension and the quadratic Casimir of the representation $R$ respectively \cite{Rusakov:1990rs}. For $SU(2)$, the representation $R$ is characterized by a positive half-integer $l$. This yields
\begin{equation}
    d_R=2l+1 \qquad \text{and} \qquad C_2(R)=l(l+1).
\end{equation}
Therefore, the partition function takes the form
\begin{equation}
    Z_{\text{YM}}(A)=\sum_{m=0}^\infty (m+1)^2 \exp \big( -\frac{e^2}{4}A((m+1)^2-1)  \big) \label{SU2partition}
\end{equation}
where $l=m/2$.

Our purpose in this section is to check whether our approach agrees to the known result by re-obtaining the partition function (\ref{SU2partition}) using our effective $SU(2)$ BF theory. To do this, we need to be more careful in integrating out the complex $b$ field in (\ref{integrate out b}) as one may notice that $\varphi^2$ in the determinant (\ref{su2 det}) will apparently get cancelled out by the Jacobian of the measure $D\phi=\varphi^2 d\varphi d\Omega$ with $\Omega$ denoting a direction of the scalar field. If the previous statement were true, we would not get the prefactor in the formula (\ref{SU2partition}). This implies that the cancellation needs to be partial. It is due to the difference in the degrees of freedom between the scalar field and the vector field.

To put it into clearer perspective, let us evaluate the $SU(2)$ partition function, i.e.
\begin{equation}
    Z=\frac{1}{\text{Vol}}\int D\phi D\chi D\bar\chi Db D\bar{b} \exp\big(-S[\phi,\chi,\bar{\chi}]-S[\phi,b,\bar{b}]\big) \label{SU(2) partition function start}
\end{equation}
where 
\begin{equation}
    S[\phi,\chi,\bar{\chi}]=2\int_\mathcal{M} d^2z \hat\phi_A(\partial\phi^A\bar{\chi}-\bar{\partial}\phi^A\chi) 
\end{equation}
and $S[\phi,b,\bar{b}]$ is expressed as (\ref{action complex field}). We then expand all the fields in terms of eigenfunctions of the scalar Laplacian, 
\begin{equation}
    \nabla^2u_\lambda=\lambda u_\lambda.
\end{equation}
Therefore, the expression for the real scalar field $\varphi$ is 
\begin{equation}
    \varphi=\sum_{\lambda\neq 0} c_\lambda u_\lambda+ \varphi_0 \label{phi basis expansion}
\end{equation}    
where the zero mode term $\varphi_0=c_0u_0$ and those for the complex vector fields are
\begin{align}
    b^\alpha&=\sum_{\lambda\neq 0} e^\alpha_\lambda \partial u_\lambda ,  &\bar{b}^\alpha&=\sum_{\lambda\neq 0} \bar{e}^\alpha_\lambda \bar\partial u_\lambda \label{b basis expansion} \\
    \chi&=\sum_{\lambda\neq 0} f_\lambda \partial u_\lambda,  &\bar{\chi}&=\sum_{\lambda\neq 0} \bar{f}_\lambda \bar\partial u_\lambda. \label{chi basis expansion}
\end{align}
 Note that there is no zero mode expansion for the vector fields as $\partial u_0=0$ and $u_\lambda$ forms a complete set of orthonormal basis satisfying
\begin{equation}
    \int d^2\xi \sqrt{g} u_\lambda(\xi) u_{\lambda'}(\xi)=\delta_{\lambda\lambda'} \qquad \text{and} \qquad \sqrt{g}\sum_\lambda u_\lambda(\xi) u_{\lambda}(\xi')=\delta^{(2)}(\xi-\xi'). \label{completeness u}
\end{equation}

Now, let first take a look at the integral 
\begin{equation}
    \int D\chi D\bar{\chi} \exp\big( -S[\phi,\chi,\bar{\chi}]\big). \label{constraint integral}
\end{equation}
By using the basis expansions, the integral (\ref{constraint integral}) takes the form
\begin{equation}
    \int|J_1| \prod_\lambda df_\lambda d\bar{f}_\lambda  \exp\bigg( 2\int d^2z \sum_{\lambda,\lambda'}c_\lambda( \partial u_\lambda \bar{\partial}u_{\lambda'}\bar{f}_{\lambda'}-\bar{\partial}u_\lambda \partial u_{\lambda'}f_{\lambda'}) \bigg) \label{constraint integral 2}
\end{equation}
where $J_1$ is the Jacobian determinant when changing  variables from $\chi$ and $\bar{\chi}$ to $f_\lambda$ and $\bar{f}_\lambda$. Therefore, it can be computed by
\begin{equation}
    J_1=\det\bigg( \frac{\delta(\chi,\bar\chi)}{\delta(f_\lambda,\bar{f}_\lambda)} \bigg) \equiv \det(M)
\end{equation}
The determinant of the matrix can be evaluated from the relation
\begin{equation}
    \text{det}(M)=\sqrt{\text{det}(M^\dag M)}. \label{matrix determinant}
\end{equation}
According to (\ref{chi basis expansion}), $\frac{\delta\chi(z)}{\delta f_\lambda}=\partial u_\lambda(z)$ and $\frac{\delta\bar\chi(z)}{\delta f_\lambda}=\bar\partial u_\lambda(z)$. Therefore,
\begin{align}
J_1&=\sqrt{\text{det}
\begin{pmatrix}
 \int d^2z \bar{\partial} u_\lambda \partial u_{\lambda'}&0 \\ 0& \int d^2z \bar{\partial} u_\lambda \partial u_{\lambda'}
\end{pmatrix}} =\prod_\lambda \lambda,
\end{align}
where (\ref{completeness u}) was utilised to obtain the last expression and the product is over the non-zero eigenvalues.

When applying the completeness relation (\ref{completeness u}) to the exponent of (\ref{constraint integral 2}), it is not hard to see that the integral becomes
\begin{align}
    \int \prod_\lambda &(-2i\lambda) d(\text{Re}(f_\lambda)) d(\text{Im}(f_\lambda))  \exp\big( 4ic_\lambda \lambda \text{Im}(f_\lambda)\big) \nonumber \\
    &= \int \prod_\lambda d(\text{Re}(f_\lambda)) (-4\pi i \lambda) \delta(4c_\lambda \lambda) \nonumber \\
     &= \prod_\lambda \text{Vol}(\text{Re}(f_\lambda)) (-\pi i  \delta(c_\lambda)). \label{constraint integral 3}
\end{align}
Similar to the expression (\ref{SU2 constraint}), the above term provides a constraint on the theory via the Dirac delta function $\delta(c_\lambda)$ requiring the modulus of the scalar field $\varphi$ to be constant, i.e. $\varphi=\varphi_0$, throughout the space.

The volume of the real number, Vol(Re($f_\lambda$)), can be cancelled with the volume of the gauge symmetry in (\ref{SU(2) partition function start}). To see this, let us apply a particular choice of  gauge-fixing to our calculation. We consider a gauge condition that makes the direction of the scalar field, $\hat{\phi}$, constant everywhere except for an infinitesimal region. After this gauge has been applied, there is left the residual gauge symmetry which does not alter the direction $\hat{\phi}$. 

Expanding an infinitesimal gauge transformation parameter $\omega$ as 
\begin{equation}
    \omega=\omega^\phi \hat{\phi}+\omega^+\hat{E}_+ +\omega^-\hat{E}_-
\end{equation}
where all components are real, the gauge transformation of the scalar field (\ref{gauge trans}) implies that the residual symmetry has $\omega^\pm=0$. Now, let us investigate the effect of this residual symmetry on the gauge field $\mathcal{A}$ where $\mathcal{A}$ takes the form 
\begin{equation}
    \mathcal{A}=\chi\hat{\phi}+b^+\hat{E}_++b^-\hat{E}_-. \label{gauge field expansion}
\end{equation}
According to (\ref{gauge trans}), a variation of the gauge field with respect to the residual gauge transformation is 
\begin{align}
    \delta_\omega \mathcal{A}&=\partial \omega+g[\mathcal{A},\omega] \nonumber \\
    &=-i\partial\omega^\phi \hat{R}-i\omega^\phi\partial \hat{R}+g\omega^\phi\bigg(\frac{1}{\sqrt{2}}(b^+-b^-) \hat{B}_1+\frac{i}{\sqrt{2}}(b^++b^-)\hat{B}_2   \bigg) \label{gauge variation}
\end{align}
where we have re-defined the bases to be 
\begin{equation}
    \hat{R}=i\hat{\phi}, \qquad \hat{B}_1=\frac{1}{\sqrt{2}}(\hat{E}_++\hat{E}_-), \qquad \hat{B}_2=\frac{-i}{\sqrt{2}}(\hat{E}_+-\hat{E}_-).
\end{equation}
Note that these bases resemble a set of ordinary unit vectors in three-dimensional sphere in which they obey the following algebras;
\begin{equation}
    [\hat{R},\hat{B}_1]=\hat{B}_2, \qquad [\hat{B}_2,\hat{R}]=\hat{B}_1, \qquad [\hat{B}_1,\hat{B}_2]=\hat{R}. \label{basis sphere algebra}
\end{equation}
As a result, if the sphere is characterised by the usual polar angle $\alpha$ and azimuthal angle $\theta$, the variation (\ref{gauge variation}) becomes
\begin{align}
    \delta_\omega \mathcal{A}=&-i\partial\omega^\phi \hat{R}+\omega^\phi\bigg( i\frac{\partial\alpha}{\partial z}\sin{\theta}+\frac{g}{\sqrt{2}}(b^+-b^-) \bigg)\hat{B}_1\nonumber \\
    &-i\omega^\phi\bigg( \frac{\partial\theta}{\partial z}-\frac{g}{\sqrt{2}}(b^++b^-) \bigg)\hat{B}_2.
\end{align}
Comparing the result to the actual variation of the gauge field (\ref{gauge field expansion}), it implies that a variation of the field $\chi$ is in the residual gauge orbit when it is real. Remember that the variation of the field $\chi$ is equivalent to that of the function $f_\lambda$ according to (\ref{chi basis expansion})). Consequently, Vol(Re($f_\lambda$)) is the residual gauge volume as claimed.

Moving on to the next integral to consider, the Gaussian functional integral of the vector fields $b^\alpha$  in the partition function (\ref{SU(2) partition function start}) can be written in terms of  scalar functions $e_\lambda$ and $\bar{e}_\lambda$ according to the Laplacian eigenfunction expansion (\ref{b basis expansion}) as
\begin{align}
   |J_2|\int  \prod_\lambda de_\lambda d\bar{e}_\lambda e^{-S[e,\bar{e}]} \label{Gaussian e}
\end{align}
where $J_2$ is the Jacobian determinant resulted from the change of variables from $b$ and $\bar{b}$ into  $e$ and $\bar{e}$. The action $S[e,\bar{e}]$ is defined as
\begin{align}
    S[e,\bar{e}]=&\int_\mathcal{M} d^2z \ \Bigg(-2ig\varphi_0\sum_{\lambda,\lambda'}\bar{e}^\alpha_\lambda\epsilon_{\alpha\beta} e^\beta_{\lambda'}\bar{\partial}u_\lambda\partial u_{\lambda'} \nonumber \\  
    &+2\varphi_0\bar\partial\hat{\phi}_A \hat{E}^A_\alpha \sum_\lambda e^\alpha_\lambda \partial u_\lambda -2\varphi_0\partial\hat{\phi}_A \hat{E}^A_\alpha \sum_\lambda \bar{e}^\alpha_\lambda \bar{\partial} u_\lambda \Bigg). \label{action complex e field}
\end{align}
To obtain the above action, the constraint (\ref{constraint integral 3}) is applied making the length of $\phi$ constant. The first term of the action (\ref{action complex e field}) vanishes when $\lambda\neq \lambda'$ due to the completeness relation (\ref{completeness u}) which yields
\begin{align}
        S[e,\bar{e}]=& 2ig\varphi_0\sum_\lambda\lambda \bar{e}^\alpha_\lambda\epsilon_{\alpha\beta} e^\beta_{\lambda}\nonumber \\  
    &+2\varphi_0 \int d^2z \Big(\bar\partial\hat{\phi}_A \hat{E}^A_\alpha \sum_\lambda e^\alpha_\lambda \partial u_\lambda -\partial\hat{\phi}_A \hat{E}^A_\alpha \sum_\lambda \bar{e}^\alpha_\lambda \bar{\partial} u_\lambda \Big). \label{action complex e field2}
\end{align}

The Jacobian determinant $J_2$ is
\begin{equation}
    J_2=\text{det}\bigg( \frac{\delta(b(z),\bar{b}(z))}{\delta(e_\lambda,\bar{e}_\lambda)} \bigg).
\end{equation}
By using the relation (\ref{matrix determinant}) and fact that $\frac{\delta b^\alpha(z)}{\delta e^\beta_\lambda}=\delta^\alpha_\beta \partial u_\lambda(z)$ and $\frac{\delta \bar{b}^\alpha(z)}{\delta \bar{e}^\beta_\lambda}=\delta^\alpha_\beta\bar{\partial} u_\lambda(z)$, one can obtain
\begin{align}
J_2&=\sqrt{\text{det}
\begin{pmatrix}
 \int d^2z \bar{\partial} u_\lambda \partial u_{\lambda'}\delta^{\alpha\beta}&0 \\ 0& \int d^2z \bar{\partial} u_\lambda \partial u_{\lambda'}\delta^{\alpha\beta}
\end{pmatrix}} =\prod_\lambda \lambda^2.
\end{align}
where (\ref{completeness u}) was utilised and again the product is over non-zero eigenvalues.

We can then calculate the Gaussian integral (\ref{Gaussian e}) over the complex field $e_\lambda$ and $\bar{e}_\lambda$ using (\ref{gaussian}). It becomes
\begin{equation}
   \int  \prod_\lambda \lambda^2 de_\lambda d\bar{e}_\lambda e^{-S[e,\bar{e}]}=\frac{\exp(-S_\text{eff}[\varphi_0.n])}{\prod_\lambda-(2g\varphi_0)^2} \label{Gaussian e2}
\end{equation}
where 
\begin{equation}
   S_\text{eff}[\varphi_0,n]=i\frac{\varphi_0}{g}\int d^2z \bar{\partial}\hat{\phi}\indices{_A}\partial\hat{\phi}\indices{_B}\hat{\phi}_C\epsilon\indices{^A^B^C}=4\pi ni\frac{\varphi_0}{g}. \label{eff action su2}
\end{equation}
Note that the effective action is related to the winding number $n$ as shown in (\ref{winding number}).

The last element to consider is the decomposition of the measure $D\phi$. This can be obtained by considering a small variation of the field $\phi$ as 
\begin{align}
    \delta\phi&=\delta\varphi \hat{\phi}+\varphi\delta\hat{\phi} \\
\intertext{with}    \delta\hat{\phi}&=\delta\omega^+\hat{E}_+ + \delta\omega^-\hat{E}_-
\end{align}
where $\delta\omega^\pm$ are small variations in the tangent directions. The variations $\delta\varphi$ and $\delta\omega^\pm$ can be expanded in terms of the eigenfunction $u_\lambda$ as 
\begin{align}
    \delta\varphi(\xi)&=\sum_m \delta c_m u_m(\xi) \\
    \delta\omega^\pm(\xi)&= \sum_m \delta \mu^\pm_m u_m(\xi)
\end{align}
Consequently, we can rewrite the measure as
\begin{equation}
    D\phi= |J_3| \prod_m dc_m d\mu^+_m d\mu^-_m\equiv  |J_3| \prod_m dc_m d\Omega \label{phi measure}
\end{equation}
where the Jacobian determinant can be computed by
\begin{equation}
    J_3= \det\bigg( \frac{\delta \phi^A(\xi)}{\delta(c_m,\mu^+_p,\mu^-_q)} \bigg)\equiv \det (M_{IJ}).
\end{equation}
$M_{IJ}$ is the Jacobian matrix where the row index $I\equiv A, \xi$ and the column index $J\equiv m,p,q$. Again, the relation (\ref{matrix determinant}) is used to determine the Jacobian determinant.

As $\frac{\delta \phi^A(\xi)}{\delta c_m}=\hat{\phi}^A(\xi)u_m(\xi)$ and $\frac{\delta \phi^A(\xi)}{\delta \mu^\pm_m}=\hat{E}^A_\pm(\xi)\varphi u_m(\xi)$, It is not hard to see that $M^\dag M$ is 
\begin{align}
\begin{pmatrix}
\big(\int_\xi u_m(\xi)u_{m'}(\xi)\big)_{mm'}&0&0 \\ 0&0&\big(\int_\xi\varphi^2 u_m(\xi)u_{m'}(\xi)\big)_{mm'} \\ 0&\big(\int_\xi \varphi^2u_m(\xi)u_{m'}(\xi)\big)_{mm'}&0
\end{pmatrix}
\end{align}
where $\int_\xi$ is a shorthand for $\int \sqrt{g} d^2\xi$. Note that the objects in the parentheses are the matrix elements in row $m$ and column $m'$. We can then utilise the fact that the value of $\varphi$ is the constant $\varphi_0$ throughout the space due to the constraint (\ref{constraint integral 3}). This allows us to obtain the absolute value of the Jacobian determinant as
\begin{equation}
    |J_3|=\prod_m (\varphi_0)^2. \label{J3}
\end{equation}

It is clear that the product of $(\varphi_0)^2$ in (\ref{J3}) cannot be completely cancelled by the one in (\ref{Gaussian e2}) as mentioned. The cancellation leaves a single factor of $(\varphi_0)^2$ behind. This remaining factor accounts for the pre-factor of the partition function as we shall see later.

In consequence, when substituting (\ref{constraint integral 3}), (\ref{Gaussian e2}), (\ref{phi measure}), and (\ref{J3}) into (\ref{integrate out b}), the gauge-fixed partition function  takes the form
\begin{equation}
    Z=\mathcal{N} \int dc_0  \bigg(\prod_\lambda dc_\lambda \delta(c_\lambda) \bigg)  \varphi_0^2 \sum_{n=-\infty}^\infty \exp (-S_\text{eff}[\varphi_0,n])
\end{equation}
where $S_\text{eff}[\varphi_0,n]$ is expressed in (\ref{eff action su2}).

According to (\ref{YM}), we can relate the BF theory to two-dimensional Yang-Mills theory by adding a quadratic term in the scalar field. Consequently, the partition function for 2D Yang-Mills is 
\begin{align}
    Z&=\widetilde{\mathcal{N}}\int_0^\infty \varphi_0^2 d\varphi_0   \sum_{n=-\infty}^\infty \exp (-S_\text{eff}[\varphi_0,n]-e^2\int_{S^2} d^2\xi\sqrt{g}\varphi^2) \nonumber \\
    &=\widetilde{\mathcal{N}} \int_0^\infty \varphi^2_0d\varphi_0 \sum_{n=-\infty}^\infty  \exp\big(-\frac{4\pi i}{g}n\varphi_0-e^2\varphi^2_0 A\big). \label{partition su2 YM}
\end{align}
where $A$ is the area of the sphere. The infinite sum of the Euler exponential provides a Dirac delta function. This discretises the possible values of $\varphi_0$ in the theory as
\begin{equation}
    \sum_{n=-\infty}^\infty \exp \big(-\frac{4\pi i}{g}  n\varphi_0\big)=\frac{g}{2}\delta \big(\varphi_0\text{mod}\frac{g}{2} \big).
\end{equation}
Therefore, it is not hard to see that the expression (\ref{partition su2 YM}) turns into
\begin{equation}
    Z\sim \sum_{m=1}^\infty m^2 \exp\big( -\frac{(eg)^2}{4}Am^2 \big). \label{su2 partition function result}
\end{equation}
The result (\ref{su2 partition function result}) is in agreement with the expression (\ref{SU2partition}). They differ by the factor $-1$ in the exponent which can be adjusted by a local counter term.

\section{Generalization to an Arbitrary Lie Algebra}
In this section, we would like to generalise the approach we used in section two to an  arbitrary Lie algebra. As seen in the earlier section, one of the key elements in our calculation is to expand the fields in terms of a set of suitable Lie bases. For a general Lie algebra, we will work in the Cartan-Weyl basis.

We will denote the Cartan generator $H^a$ and Weyl generator $E^\alpha$ where $a=1,\ldots, N-1$ and $\alpha$ is a root of eigenvalue equation, $ \text{ad}_{H^a}(E^\alpha)=\alpha^a E^\alpha$. The roots $\alpha$ forms a vector space $\Phi$. The generators $H^a$ and $E^\alpha$ satisfy the following algebra:
\begin{align}
    [H^a,H^b]=0, \qquad [H^a,E^\alpha]=\alpha^{(a)} E^\alpha, 
    \nonumber \\
    \text{and} \qquad [E^\alpha,E^\beta]=    \begin{cases}
      N^{\alpha\beta}E^{\alpha+\beta} & \text{if} \ \alpha+\beta\in \Phi\\
      H^\alpha & \text{if} \ \alpha+\beta=0\\
    \end{cases} 
\end{align}
where $H^\alpha$ is defined as $H^\alpha=\alpha_{a}H^a$. The Cartan generators $H^a$ are diagonal traceless matrices in the adjoint representation.

Again, we start the calculation with the action (\ref{topological field action2}) with the path integral defined by (\ref{path integral}). The calculation proceeds by expanding the fields $\phi$ and $\mathcal{A}_i$ in the Cartan-Weyl basis as
\begin{equation}
    \phi=\phi_aH^a \qquad \text{and} \qquad \mathcal{A}_i=\chi_{ia}H^a+a_{i\alpha} E^\alpha.
\end{equation}
Similar to the SU(2) case, these bases are $\xi$-dependent. The Cartan generators were chosen such that the field $\phi$ lies within their subalgebra.

To relate Lie indices $A$ with the Cartan and Weyl indices $a$ and $\alpha$, we introduce unit vectors $\hat{H}^a_A$ and $\hat{E}^\alpha_A$ in Lie vector space which are defined as $\delta^a_A$ and $\delta^\alpha_A$ respectively. As a result, the inner products among the vectors are
\begin{equation}
    \hat{H}^a_A\hat{H}^{Ab}=\eta^{ab}, \qquad \hat{E}^\alpha_A\hat{E}^{A\beta}=\eta^{\alpha\beta},\qquad  \hat{H}^a_A\hat{E}^{A\alpha}=0
\end{equation}
and the completeness relation is
\begin{equation}
     \hat{H}^A_a\hat{H}^a_B+\hat{E}^A_\alpha\hat{E}^\alpha_B=\delta^{A}_B. \label{completeness}
\end{equation}
It is not hard to write the field $\phi$ and $\mathcal{A}_i$ in terms of the unit vectors as
\begin{equation}
     \phi^A=\phi^a\hat{H}^A_a \qquad \text{and} \qquad \mathcal{A}_i^A=\chi^a_i\hat{H}^A_a+a^\alpha_i \hat{E}^A_\alpha. \label{Lie comp}
\end{equation}

Using the relations (\ref{Lie comp}), one can find the topological field  theory action (\ref{topological field action2}) as 
\begin{equation}
    S[\phi,\chi,a]=\int_{\mathcal{M}} d^2\xi \ \Bigg(ig f^{ABC} \phi\indices{_C} a^\alpha_{i} a^\beta_{j} \hat{E}_{\alpha A}\hat{E}_{\beta B}  -2(\partial_i\phi_A)a^\alpha_j\hat{E}^A_\alpha -2(\partial_i\phi_A)\chi^a_j\hat{H}^A_a\Bigg) \epsilon^{ij}.  \label{topological field action3}
\end{equation}
Notice that there is no contribution from diagonal components of $\mathcal{A}_i^A$ to the first term as the Cartan subalgebra is commutative.

To obtain the effective Lagrangian of the field $\phi$, we need to integrate out the variables $\chi^a_i$ and $a_i^\alpha$. According to the action (\ref{topological field action3}), integrating out $\chi^a_i$ would provide a constraint via the Dirac-delta function as 
\begin{align}
    \int D\chi_{ja} \ \text{exp}(2\int d^2\xi (\partial_i\phi^A)\chi_{ja}\hat{H}^a_A\epsilon^{ij} )&= \mathcal{N}\prod_{a=1}^{N-1}\delta^{(2)}((\underline{\partial}\phi^A)\hat{H}^a_A) \nonumber \\
    &= \mathcal{N}\prod_{a=1}^{N-1}\delta^{(2)}(2\text{tr}((\underline{\partial}\phi)H^a)). \label{constraint}
\end{align}
This implies that the derivative of the field $\phi$, i.e. $\partial_i\phi$, has no $H^a$ component. This provides a constraint on the theory as tr$(\phi\partial_i \phi)=0$.

This constraint (\ref{constraint}) also implies that the square of the field $\phi$, i.e. $\phi\indices{^A}\phi\indices{_A}\equiv |\phi|^2$, is constant throughout the space which is similar to what we found earlier in the SU(2) theory. Apart from that, it also implies the existence of the new invariant quantity,
\begin{equation}
    d^{ABC}\phi\indices{_A}\phi\indices{_B}\phi\indices{_C}
\end{equation}
where $d^{ABC}$ is a totally symmetric third rank tensor defined by
\begin{equation}
    d^{ABC}=2\text{tr}(\{T^A,T^B\}T^C). \label{d factor}
\end{equation}

Up to this point we have ignored a boundary in (\ref{topological field action2}). We will now consider the effect of including this term $2\int d^2\xi \partial_i(\phi\indices{_A}\mathcal{A}\indices{^A_j})\epsilon^{ij}$. It affects the constraints. To see this, let consider the case when the manifold $\mathcal{M}$ has the topology of a disk. This manifold can be mapped to the upper-half plane parameterised by Cartesian coordinates. Therefore, the boundary term takes the form 
\begin{equation}
    -2\int d^2\mathbf{x} \delta(y) \phi\indices{_A}\mathcal{A}\indices{^A_x}.
\end{equation}
By expanding the gauge field $\mathcal{A}$ as (\ref{Lie comp}), this turns the theory constraints (\ref{constraint}) into
\begin{equation}
    \prod_a \delta(2\text{tr}(\partial_x\phi)H^a)\delta(2\text{tr}(\partial_y\phi-\delta(y)\phi)H^a).
\end{equation}
This implies that the squared of the field $\phi$ is no longer constant throughout the manifold $\mathcal{M}$. There is a discontinuity of $|\phi|^2$ at the boundary in the $y$ direction as
\begin{equation}
    |\phi|^2(x,\epsilon)=3|\phi|^2(x,0).
\end{equation}

To perform the path integration with respect to the field $a^\alpha_i$, we apply the same trick we used in the previous section. We change the spacetime coordinates $\xi^1$ and $\xi^2$ into the complex coordinates $z$ and $\bar{z}$ which were previously defined in (\ref{complex coord}). Of course, this coordinate transformation modifies the field $a^\alpha_i$ into the complex field $b^\alpha$ as stated in (\ref{complex b field}).

As a result, the partition function now takes form
\begin{equation}
    Z=\frac{1}{\text{Vol}}\int D\phi\indices{_A} Db^\alpha D\bar{b}^\alpha \ \prod_{a=1}^{N-1}\delta^{(2)}(2\text{tr}((\underline{\partial}\phi)H^a))\text{exp}(-S[\phi,b,\bar{b}]) \label{partition function}
\end{equation}
where the action is expressed in the complex coordinates as
\begin{equation}
    S[\phi,b,\bar{b}]=2\int_D d^2z \Bigg( igf^{ABC}\phi\indices{_C} b^\alpha\bar{b}^\beta \hat{E}_{\alpha A}\hat{E}_{\beta B} -(\partial\phi_A\bar{b}^\alpha-\bar{\partial}\phi_Ab^\alpha)\hat{E}^A_\alpha \Bigg). \label{SUN action b field}
\end{equation}

The path integral of the complex fields $b^\alpha$ and $\bar{b}^\alpha$ resembles a Gaussian integral which can be performed using (\ref{gaussian}). By comparing (\ref{SUN action b field}) with (\ref{gaussian}), one obtains
\begin{align}
    M_{\alpha\beta}=2gif^{ABC}\phi\indices{_B}\hat{E}_{\alpha A}\hat{E}_{\beta C}, \qquad J_\alpha=-2\hat{E}^A_\alpha\partial \phi\indices{_A}, \qquad \bar{J}_\alpha=2\hat{E}^A_\alpha\bar{\partial} \phi\indices{_A}. \label{M}
\end{align}
Consequently, it is not hard to find that the effective Lagrangian with respect to the scalar field $\phi$ is 
\begin{align}
    \mathcal{L}_\text{eff}(\phi)=\frac{-i}{2g}\bar{J}_\alpha (\widetilde{M}^{-1})\indices{^\alpha_\beta} J^\beta\ = \ \frac{2i}{g}\partial\phi\indices{_A} \bar{\partial}\phi\indices{_B} \Big( \hat{E}^A_\alpha (\widetilde{M}^{-1})\indices{^\alpha_\beta} \hat{E}^B_\beta  \Big) \label{effective SUN}
\end{align}
where we used $M\indices{^\alpha_\beta}=2gi\widetilde{M}\indices{^\alpha_\beta}$.

A general expression for an inverse matrix $\widetilde{M}\indices{^\alpha_\beta}$ is
\begin{equation}
    (\widetilde{M}^{-1})\indices{^\alpha_\beta}=\frac{\text{adj}(\widetilde{M})\indices{^\alpha_\beta}}{\text{det}(\widetilde{M})} \label{inverse}
\end{equation}
where
\begin{align}
    \text{adj}(\widetilde{M})\indices{^\alpha_\beta}=& \ \delta^{\alpha j_2\ldots j_n}_{\beta i_2\ldots i_n} \  \widetilde{M}\indices{^{i_2}_{j_2}}\widetilde{M}\indices{^{i_3}_{j_3}}\ldots \widetilde{M}\indices{^{i_n}_{j_n}}, \nonumber \\ 
    \text{det}(\widetilde{M})=& \ \delta^{j_1j_2\ldots j_n}_{i_1 i_2\ldots i_n}\ \widetilde{M}\indices{^{i_1}_{j_1}}\widetilde{M}\indices{^{i_2}_{j_2}}\ldots \widetilde{M}\indices{^{i_n}_{j_n}}. \label{adj det}
\end{align}
$\delta^{j_1j_2\ldots j_n}_{i_1 i_2\ldots i_n}$ is a generalised Kronecker delta which is related to an anti-symmetrization of ordinary Kronecker deltas as
\begin{equation}
    \delta^{j_1j_2\ldots j_n}_{i_1 i_2\ldots i_n}=n!\delta^{j_1}_{[i_1}\delta^{j_2}_{i_2}\ldots\delta^{j_n}_{i_n]}. \label{generalised Kronecker delta}
\end{equation}
The integer $n$ is the number of Weyl generators. In the case of $SU(N)$, $n$ is equal to $N^2-N$.

To obtain the adjugate matrix and the matrix determinant expressed in (\ref{adj det}), the matrices $\widetilde{M}\indices{^i_j}$ are contracted with each other depending on the permutations implicit by  (\ref{generalised Kronecker delta}). For the adjugate matrix $\text{adj}(\widetilde{M})\indices{^\alpha_\beta}$, the contractions lead to two types of terms. First, the matrices $\widetilde{M}\indices{^i_j}$ are contracted in such a way that they form a new matrix with indices $\alpha$ and $\beta$. This contraction generates a chain of matrix multiplications, for instance, $\widetilde{M}\indices{^\alpha_{j_2}}\widetilde{M}\indices{^{j_2}_{j_3}}\widetilde{M}\indices{^{j_4}_{j_4}}\widetilde{M}\indices{^{j_4}_\beta}$. In this example, the matrices $\widetilde{M}\indices{^{i_2}_{j_2}}\widetilde{M}\indices{^{i_3}_{j_3}}\widetilde{M}\indices{^{i_4}_{j_4}}\widetilde{M}\indices{^{i_5}_{j_5}}$ are contracted with $\delta^\alpha_{i_2}\delta^{j_2}_{i_3}\delta^{j_3}_{i_4}\delta^{j_4}_{i_5}\delta^{j_5}_\beta$. Second, the contraction forms a trace of matrix products, i.e. tr($\widetilde{M}\cdot \widetilde{M}\ldots \widetilde{M}$). For example, when the same matrices $\widetilde{M}\indices{^{i_2}_{j_2}}\widetilde{M}\indices{^{i_3}_{j_3}}\widetilde{M}\indices{^{i_4}_{j_4}}\widetilde{M}\indices{^{i_5}_{j_5}}$ are contracted with $\delta^{j_2}_{i_3}\delta^{j_3}_{i_4}\delta^{j_4}_{i_5}\delta^{j_5}_{i_2}$. However, only the latter case contributes to the matrix determinant det($\widetilde{M}$). 

In addition, the trace term vanishes when the number of matrices $\widetilde{M}$ inside is odd. This can be seen explicitly by considering 
\begin{align}
    \text{tr}&(\widetilde{M}\cdot \widetilde{M}\ldots \widetilde{M})= \widetilde{M}\indices{^\alpha_{i_1}}\widetilde{M}\indices{^{i_1}_{i_2}}\ldots \widetilde{M}\indices{^{i_{k-2}}_{i_{k-1}}}\widetilde{M}\indices{^{i_{k-1}}_\alpha} \nonumber \\
    &= f^{A_1B_1C_1}\phi\indices{_{B_1}}\eta\indices{_{C_1}_{A_2}}f^{A_2B_2C_2}\phi\indices{_{B_2}}\eta\indices{_{C_2}_{A_3}}\cdot\ldots\cdot f^{A_kB_kC_k
    }\phi\indices{_{B_k}}\eta\indices{_{C_k}_{A_1}}.
\end{align}
We used the completeness relation (\ref{completeness}) to obtain the last line. When we swap the first and the third indices of each structure constant $f^{ABC}$, it gives an extra $(-1)$ to the last line so the whole expression vanishes. 

The calculation of the inverse matrix (\ref{inverse}) involves a lot of contractions  corresponding to chains of matrix multiplications. To facilitate the calculation, it is sensible to develop a set of diagrams to represent them. These diagrams are presented in the next section.

\section{Diagrammatic Representation of the Inverse Matrix $\widetilde{M}$}
\begin{figure}[t]
\begin{center}
\begin{tabular}{ c l }
$\vcenter{\hbox{\begin{tikzpicture}
\def\r{0.5};
\draw (0,0) circle [radius=\r];
\draw (0,-\r)--(0,\r);
\node at (0:0.5*\r) {\footnotesize{$j$}};
\node at (0:-0.5*\r) {\footnotesize{$i$}};
\end{tikzpicture}}}$ & $\equiv \ \widetilde{M}\indices{^i_j} \qquad = \hat{E}^{i}_A(f^{ABC}\phi\indices{_B})\hat{E}_{jC}$  \\[0.5cm]
 $\vcenter{\hbox{
\begin{tikzpicture}
\draw (-1,0)--(1,0);
\node[left] at (-1,0) {\footnotesize{$i$}};
\node[right] at (1,0) {\footnotesize{$j$}};
\end{tikzpicture}}}$ & $\equiv \ \delta^i_j.$
\end{tabular}
\end{center}
    \caption{Diagrammatic representation for matrix element $\widetilde{\mathcal{M}}$ and Kronecker delta}
    \label{vertex line}
\end{figure}

According to the previous section, the inverse of the matrix $\widetilde{M}$ is an essential ingredient of the $SU(N)$ effective Lagrangian (\ref{effective SUN}). To compute this object, the relation (\ref{inverse}) is used. However, this is complicated by the large number of terms.

For this reason, we would like to develop a set of diagrams to capture the contractions between matrix elements $\widetilde{M}\indices{^\alpha_\beta}$and Kronecker deltas  $\delta^i_j$. We represent these two objects as the vertices and lines shown in figure \ref{vertex line}.

Based on this diagrammatic representation, matrix multiplication is represented by vertices connecting by a line. Note that no more than two lines are allowed to be connected to each vertex. This fact implies that a diagram involved in the calculation is either a strand or a loop which corresponds to a chain of matrix multiplications and its trace respectively. Just for clarification, we show some examples for a loop diagram and a strand diagram as well as their corresponding matrix representations in figure \ref{diagram example}.

\begin{figure}[t]
\begin{center}
\addtolength{\tabcolsep}{-5pt}
\begin{tabular}{ c l l }
$\vcenter{\hbox{\begin{tikzpicture}
\def\r{0.3};
\def\l{0.4};
\draw (0,0) circle [radius=\r];
\draw (0,-\r)--(0,\r);
\draw (\r,0)--(\r+\l,0) (-\r,0)--(-\r-\l,0);
\draw (\l+2*\r,0) circle [radius=\r];
\draw (\l+2*\r,-\r)--(\l+2*\r,\r);
\draw (-\l-2*\r,0) circle [radius=\r];
\draw (-\l-2*\r,-\r)--(-\l-2*\r,\r);
\node[above right] at (0:\l+2.7*\r) {\footnotesize{$\beta$}};
\node[above left] at (0:-\l-2.7*\r) {\footnotesize{$\alpha$}};
\end{tikzpicture}}}$ & $= \ \widetilde{M}\indices{^\alpha_i}\widetilde{M}\indices{^i_j}\widetilde{M}\indices{^j_\beta}$ & =\raisebox{-0.3 cm}{\hbox{$\begin{aligned} \hat{E}^{\alpha}_A(&f\indices{^A^B^C}\phi\indices{_B}\eta\indices{_C_D}f\indices{^D^E^F}\phi\indices{_E} \\ & \times\eta\indices{_F_G}f\indices{^G^H^I}\phi\indices{_H})\hat{E}_{\beta I} \end{aligned}$}} \\[0.5cm]
 $\vcenter{\hbox{
\begin{tikzpicture}
\def\R{1};
\def\r{0.3};
\draw (0,0) circle [radius=\R];
\draw[fill=white] (45:\R) circle [radius=\r];
\draw[fill=white] (135:\R) circle [radius=\r];
\draw[fill=white] (225:\R) circle [radius=\r];
\draw[fill=white] (315:\R) circle [radius=\r];
\draw ([yshift=-\r cm]45:\R)--([yshift=\r cm]45:\R);
\draw ([yshift=-\r cm]135:\R)--([yshift=\r cm]135:\R);
\draw ([yshift=-\r cm]225:\R)--([yshift=\r cm]225:\R);
\draw ([yshift=-\r cm]315:\R)--([yshift=\r cm]315:\R);
\end{tikzpicture}}}$ & $= \ \widetilde{M}\indices{^i_j}\widetilde{M}\indices{^j_k}\widetilde{M}\indices{^k_l}\widetilde{M}\indices{^l_i}$ & 
    = \raisebox{-0.3 cm}{\hbox{$\begin{aligned}
&f\indices{^A^B^C}\phi\indices{_B}\eta\indices{_C_D}f\indices{^D^E^F}\phi\indices{_E}\eta\indices{_F_G} \\ &\times f\indices{^G^H^I}\phi\indices{_H}\eta\indices{_I_J}f\indices{^J^K^L}\phi\indices{_K}\eta\indices{_L_A}    
\end{aligned}$}}
\end{tabular}
\end{center}
    \caption{Examples for a strand and loop diagram representing certain matrix multiplications}
    \label{diagram example}
\end{figure}

According to (\ref{adj det}), the adjugate matrix, adj$(\widetilde{M})\indices{^\alpha_\beta}$, can be expressed diagrammatically as a summation of all possible products between a strand diagram and loop diagrams. The diagram includes  $n-1$ vertices in total where $n=N^2-N$ for $SU(N)$ ($n$ is always even for $N\geq 2$). In order to obtain all possible combinations of a strand and loops without overcounting, we can start by listing all possible strand diagrams which simply are the strand with different numbers of vertices ranging from $1$ to $n-1$. Then, for each strand, loop diagrams can be created using the remaining vertices. Therefore, we can expand the adjugate matrix as
\begin{align}
&\text{adj}(\widetilde{M})\indices{^\alpha_\beta}=(-1)^{n-1}\Bigg\{(n-1)! \vcenter{\hbox{\begin{tikzpicture}
\def\r{0.25};
\def\l{0.4};
\draw (0,0) circle [radius=\r];
\draw (0,-\r)--(0,\r);
\draw (\r,0)--(\r+\l,0) (-\r,0)--(-\r-\l,0);
\draw (-\l-2*\r,0) circle [radius=\r];
\draw (-\l-2*\r,-\r)--(-\l-2*\r,\r);
\node[above left] at (-\l-2.7*\r,0) {\footnotesize{$\alpha$}};
\node[below] at (\r+1.5*\l,0) {\ldots};
\draw (\r+2*\l,0)--(\r+3*\l,0);
\draw (2*\r+3*\l,0) circle [radius=\r];
\draw (2*\r+3*\l,-\r)--(2*\r+3*\l,\r);
\draw (3*\r+3*\l,0)--(3*\r+4*\l,0);
\draw (4*\r+4*\l,0) circle [radius=\r];
\draw (4*\r+4*\l,-\r)--(4*\r+4*\l,\r);
\node[above right] at (4*\l+4.7*\r,0) {\footnotesize{$\beta$}};
\draw [decorate,decoration={brace,amplitude=8pt,mirror,raise=10pt}]
(-3*\r-\l,0) -- (5*\r+4*\l,0) node [black,midway,yshift=-1cm] 
{\footnotesize $(n-1) \ \text{terms}$};
\end{tikzpicture}}} \nonumber \\
&-(n-3)! \vcenter{\hbox{\begin{tikzpicture}
\def\r{0.25};
\def\l{0.4};
\draw (0,0) circle [radius=\r];
\draw (0,-\r)--(0,\r);
\draw (\r,0)--(\r+\l,0) (-\r,0)--(-\r-\l,0);
\draw (-\l-2*\r,0) circle [radius=\r];
\draw (-\l-2*\r,-\r)--(-\l-2*\r,\r);
\node[above left] at (-\l-2.7*\r,0) {\footnotesize{$\alpha$}};
\node[below] at (\r+1.5*\l,0) {\ldots};
\draw (\r+2*\l,0)--(\r+3*\l,0);
\draw (2*\r+3*\l,0) circle [radius=\r];
\draw (2*\r+3*\l,-\r)--(2*\r+3*\l,\r);
\draw (3*\r+3*\l,0)--(3*\r+4*\l,0);
\draw (4*\r+4*\l,0) circle [radius=\r];
\draw (4*\r+4*\l,-\r)--(4*\r+4*\l,\r);
\node[above right] at (4*\l+4.7*\r,0) {\footnotesize{$\beta$}};
\draw [decorate,decoration={brace,amplitude=8pt,mirror,raise=10pt}]
(-3*\r-\l,0) -- (5*\r+4*\l,0) node [black,midway,yshift=-1cm] 
{\footnotesize $(n-3) \ \text{terms}$};
\end{tikzpicture}}} \times \binom{n-1}{2} \vcenter{\hbox{
\begin{tikzpicture}
\def\R{.4};
\def\r{0.2};
\draw (0,0) circle [radius=\R];
\draw[fill=white] (90:\R) circle [radius=\r];
\draw[fill=white] (-90:\R) circle [radius=\r];
\draw ([yshift=-\r cm]90:\R)--([yshift=\r cm]90:\R);
\draw ([yshift=-\r cm]-90:\R)--([yshift=\r cm]-90:\R);
\end{tikzpicture}}} \nonumber \\
&-(n-5)! \vcenter{\hbox{\begin{tikzpicture}
\def\r{0.25};
\def\l{0.4};
\draw (0,0) circle [radius=\r];
\draw (0,-\r)--(0,\r);
\draw (\r,0)--(\r+\l,0) (-\r,0)--(-\r-\l,0);
\draw (-\l-2*\r,0) circle [radius=\r];
\draw (-\l-2*\r,-\r)--(-\l-2*\r,\r);
\node[above left] at (-\l-2.7*\r,0) {\footnotesize{$\alpha$}};
\node[below] at (\r+1.5*\l,0) {\ldots};
\draw (\r+2*\l,0)--(\r+3*\l,0);
\draw (2*\r+3*\l,0) circle [radius=\r];
\draw (2*\r+3*\l,-\r)--(2*\r+3*\l,\r);
\draw (3*\r+3*\l,0)--(3*\r+4*\l,0);
\draw (4*\r+4*\l,0) circle [radius=\r];
\draw (4*\r+4*\l,-\r)--(4*\r+4*\l,\r);
\node[above right] at (4*\l+4.7*\r,0) {\footnotesize{$\beta$}};
\draw [decorate,decoration={brace,amplitude=8pt,mirror,raise=10pt}]
(-3*\r-\l,0) -- (5*\r+4*\l,0) node [black,midway,yshift=-1cm] 
{\footnotesize $(n-5) \ \text{terms}$};
\end{tikzpicture}}} \times \binom{n-1}{4} \Bigg[ 3!
\vcenter{\hbox{
\begin{tikzpicture}
\def\R{.5};
\def\r{0.2};
\draw (0,0) circle [radius=\R];
\draw[fill=white] (45:\R) circle [radius=\r];
\draw[fill=white] (135:\R) circle [radius=\r];
\draw[fill=white] (225:\R) circle [radius=\r];
\draw[fill=white] (315:\R) circle [radius=\r];
\draw ([yshift=-\r cm]45:\R)--([yshift=\r cm]45:\R);
\draw ([yshift=-\r cm]135:\R)--([yshift=\r cm]135:\R);
\draw ([yshift=-\r cm]225:\R)--([yshift=\r cm]225:\R);
\draw ([yshift=-\r cm]315:\R)--([yshift=\r cm]315:\R);
\end{tikzpicture}}}
-\frac{\binom{4}{2}\binom{2}{2}}{2!}
\vcenter{\hbox{
\begin{tikzpicture}
\def\R{.4};
\def\r{0.2};
\draw (0,0) circle [radius=\R];
\draw[fill=white] (90:\R) circle [radius=\r];
\draw[fill=white] (-90:\R) circle [radius=\r];
\draw ([yshift=-\r cm]90:\R)--([yshift=\r cm]90:\R);
\draw ([yshift=-\r cm]-90:\R)--([yshift=\r cm]-90:\R);
\end{tikzpicture}}}
\vcenter{\hbox{
\begin{tikzpicture}
\def\R{.4};
\def\r{0.2};
\draw (0,0) circle [radius=\R];
\draw[fill=white] (90:\R) circle [radius=\r];
\draw[fill=white] (-90:\R) circle [radius=\r];
\draw ([yshift=-\r cm]90:\R)--([yshift=\r cm]90:\R);
\draw ([yshift=-\r cm]-90:\R)--([yshift=\r cm]-90:\R);
\end{tikzpicture}}} \Bigg] \nonumber \\
&-\ldots - 1! \raisebox{-0.1 cm}{\hbox{
\begin{tikzpicture}
\def\r{0.25};
\draw (0,0) circle [radius=\r];
\draw ([yshift=-\r cm]0,0)--([yshift=\r cm]0,0);
\node[above left] at (-0.7*\r,0) {\footnotesize{$\alpha$}};
\node[above right] at (0.7*\r,0) {\footnotesize{$\beta$}};
\end{tikzpicture}}}
\times \binom{n-1}{n-2}\Bigg[  (n-3)!
\vcenter{\hbox{
\begin{tikzpicture}
\def\R{.7};
\def\r{0.2};
\def\s{0.1cm};
\draw ([xshift=\s]-60:\R) arc (-60:60:\R);
\draw ([xshift=-\s]120:\R) arc (120:240:\R);
\draw[fill=white] ([xshift=\s]0:\R) circle [radius=\r];
\draw[fill=white] ([xshift=\s]60:\R) circle [radius=\r];
\draw[fill=white] ([xshift=-\s]120:\R) circle [radius=\r];
\draw[fill=white] ([xshift=-\s]180:\R) circle [radius=\r];
\draw[fill=white] ([xshift=-\s]240:\R) circle [radius=\r];
\draw[fill=white] ([xshift=\s]300:\R) circle [radius=\r];
\draw ([xshift=\s,yshift=-\r cm]0:\R)--([xshift=\s,yshift=\r cm]0:\R);
\draw ([xshift=\s,yshift=-\r cm]60:\R)--([xshift=\s,yshift=\r cm]60:\R);
\draw ([xshift=-\s,yshift=-\r cm]120:\R)--([xshift=-\s,yshift=\r cm]120:\R);
\draw ([xshift=-\s,yshift=-\r cm]180:\R)--([xshift=-\s,yshift=\r cm]180:\R);
\draw ([xshift=-\s,yshift=-\r cm]240:\R)--([xshift=-\s,yshift=\r cm]240:\R);
\draw ([xshift=\s,yshift=-\r cm]300:\R)--([xshift=\s,yshift=\r cm]300:\R);
\node at (270:\R) {\tiny{\dots}};
\node at (90:\R) {\tiny{\dots}};
\draw [decorate,decoration={brace,amplitude=8pt,mirror,raise=5pt}]
(-\R-\r,-\R) -- (\R+\r,-\R) node [black,midway,yshift=-0.8cm] 
{\footnotesize $(n-2) \ \text{terms}$}; 
\end{tikzpicture}}} \nonumber \\ 
& \hspace{5.4cm} +\ldots+ (-1)^{\frac{n-2}{2}-1}
\vcenter{\hbox{
\begin{tikzpicture}
\def\R{.4cm};
\def\r{0.2};
\def\s{1cm};
\draw (0,0) circle [radius=\R];
\draw[fill=white] (90:\R) circle [radius=\r];
\draw[fill=white] (-90:\R) circle [radius=\r];
\draw ([yshift=-\r cm]90:\R)--([yshift=\r cm]90:\R);
\draw ([yshift=-\r cm]-90:\R)--([yshift=\r cm]-90:\R);
\draw ([xshift=\s]0,0) circle [radius=\R];
\draw[fill=white] ([xshift=\s]90:\R) circle [radius=\r];
\draw[fill=white] ([xshift=\s]-90:\R) circle [radius=\r];
\draw ([xshift=\s,yshift=-\r cm]90:\R)--([xshift=\s,yshift=\r cm]90:\R);
\draw ([xshift=\s,yshift=-\r cm]-90:\R)--([xshift=\s,yshift=\r cm]-90:\R);
\node at (1.5*\s+0.75*\R,0) {\ldots};
\draw ([xshift=2*\s+1.5*\R]0,0) circle [radius=\R];
\draw[fill=white] ([xshift=2*\s+1.5*\R]90:\R) circle [radius=\r];
\draw[fill=white] ([xshift=2*\s+1.5*\R]-90:\R) circle [radius=\r];
\draw ([xshift=2*\s+1.5*\R,yshift=-\r cm]90:\R)--([xshift=2*\s+1.5*\R,yshift=\r cm]90:\R);
\draw ([xshift=2*\s+1.5*\R,yshift=-\r cm]-90:\R)--([xshift=2*\s+1.5*\R,yshift=\r cm]-90:\R);
\draw [decorate,decoration={brace,amplitude=8pt,mirror,raise=8pt}]
(-\R,-\R) -- (2*\s+2.5*\R,-\R) node [black,midway,yshift=-0.8cm] 
{\footnotesize $(n-2)/2 \ \text{loops}$}; 
\end{tikzpicture}}}
\ \Bigg] \Bigg\}. \label{adjugate}
\end{align}
There is no contribution from loops with odd vertices as they are zero as discussed previously. The minus sign factor comes from an antisymmetric permutation of the generalised Kronecker delta. Each time the diagram collapses to form smaller loops, an extra (-1) appears which corresponds to an odd permutation of the lower indices of the Kronecker delta in (\ref{generalised Kronecker delta}). The numbers in front of the diagrams count the multiplicities.

One can also see that the indices $\alpha$ and $\beta$ from the adj$(\widetilde{M})\indices{^\alpha_\beta}$ are embedded at the ends of the strands corresponding to the basis $\hat{E}^{A\alpha}$. Consequently, we can always factor out these bases to write the adjugate matrix as 
\begin{equation}
    \text{adj}(\widetilde{M})\indices{^\alpha_\beta}=\hat{E}^{\alpha}_A\Theta^{AB}\hat{E}_{\beta B}
\end{equation}
or equivalently $\Theta^{AB}=\hat{E}^{A}_\alpha(\text{adj}(\widetilde{M})\indices{^\alpha_\beta})\hat{E}^{\beta B}$. Due to the above relation, it is not hard to see that the effective Lagrangian takes the form
\begin{equation}
\mathcal{L}_\text{eff}(\phi)= \ \frac{2i}{g} \frac{1}{\text{det}(\widetilde{M})}\partial\phi\indices{_A} \Theta^{AB} \bar{\partial}\phi\indices{_B}.  \label{effective}
\end{equation}

Unlike the adjugate matrix, only loop diagrams contribute to the matrix determinant det$(\widetilde{M})$. There are $n$ vertices involve in the expression of the matrix determinant. Det$(\widetilde{M})$ is expressed as the sum over all product of loops. To obtain these, we can start with the biggest loop of $n$ vertices and then cut it down to form smaller loops. The expression for det$(\widetilde{M})$ is shown in the equation (\ref{detM}). To avoid overcounting, all diagrams in the squared brackets contain the same number for fewer vertices than the loop in front of the bracket. Therefore, the general expression for the determinant is

\begin{align}
&\det(
\widetilde{M})=(-1)^{n-1} \Bigg\{ (n-1)! \vcenter{\hbox{
\begin{tikzpicture}
\def\R{.7};
\def\r{0.2};
\def\s{0.1cm};
\draw ([xshift=\s]-60:\R) arc (-60:60:\R);
\draw ([xshift=-\s]120:\R) arc (120:240:\R);
\draw[fill=white] ([xshift=\s]0:\R) circle [radius=\r];
\draw[fill=white] ([xshift=\s]60:\R) circle [radius=\r];
\draw[fill=white] ([xshift=-\s]120:\R) circle [radius=\r];
\draw[fill=white] ([xshift=-\s]180:\R) circle [radius=\r];
\draw[fill=white] ([xshift=-\s]240:\R) circle [radius=\r];
\draw[fill=white] ([xshift=\s]300:\R) circle [radius=\r];
\draw ([xshift=\s,yshift=-\r cm]0:\R)--([xshift=\s,yshift=\r cm]0:\R);
\draw ([xshift=\s,yshift=-\r cm]60:\R)--([xshift=\s,yshift=\r cm]60:\R);
\draw ([xshift=-\s,yshift=-\r cm]120:\R)--([xshift=-\s,yshift=\r cm]120:\R);
\draw ([xshift=-\s,yshift=-\r cm]180:\R)--([xshift=-\s,yshift=\r cm]180:\R);
\draw ([xshift=-\s,yshift=-\r cm]240:\R)--([xshift=-\s,yshift=\r cm]240:\R);
\draw ([xshift=\s,yshift=-\r cm]300:\R)--([xshift=\s,yshift=\r cm]300:\R);
\node at (270:\R) {\tiny{\dots}};
\node at (90:\R) {\tiny{\dots}};
\draw [decorate,decoration={brace,amplitude=8pt,mirror,raise=5pt}]
(-\R-\r,-\R) -- (\R+\r,-\R) node [black,midway,yshift=-0.8cm] 
{\footnotesize $n \ \text{terms}$}; 
\end{tikzpicture}}} 
-\binom{n}{2}(n-3)! \vcenter{\hbox{
\begin{tikzpicture}
\def\R{.7};
\def\r{0.2};
\def\s{0.1cm};
\draw ([xshift=\s]-60:\R) arc (-60:60:\R);
\draw ([xshift=-\s]120:\R) arc (120:240:\R);
\draw[fill=white] ([xshift=\s]0:\R) circle [radius=\r];
\draw[fill=white] ([xshift=\s]60:\R) circle [radius=\r];
\draw[fill=white] ([xshift=-\s]120:\R) circle [radius=\r];
\draw[fill=white] ([xshift=-\s]180:\R) circle [radius=\r];
\draw[fill=white] ([xshift=-\s]240:\R) circle [radius=\r];
\draw[fill=white] ([xshift=\s]300:\R) circle [radius=\r];
\draw ([xshift=\s,yshift=-\r cm]0:\R)--([xshift=\s,yshift=\r cm]0:\R);
\draw ([xshift=\s,yshift=-\r cm]60:\R)--([xshift=\s,yshift=\r cm]60:\R);
\draw ([xshift=-\s,yshift=-\r cm]120:\R)--([xshift=-\s,yshift=\r cm]120:\R);
\draw ([xshift=-\s,yshift=-\r cm]180:\R)--([xshift=-\s,yshift=\r cm]180:\R);
\draw ([xshift=-\s,yshift=-\r cm]240:\R)--([xshift=-\s,yshift=\r cm]240:\R);
\draw ([xshift=\s,yshift=-\r cm]300:\R)--([xshift=\s,yshift=\r cm]300:\R);
\node at (270:\R) {\tiny{\dots}};
\node at (90:\R) {\tiny{\dots}};
\draw [decorate,decoration={brace,amplitude=8pt,mirror,raise=5pt}]
(-\R-\r,-\R) -- (\R+\r,-\R) node [black,midway,yshift=-0.8cm] 
{\footnotesize $(n-2) \ \text{terms}$}; 
\end{tikzpicture}}} \times \Bigg[
\vcenter{\hbox{
\begin{tikzpicture}
\def\R{.4};
\def\r{0.2};
\draw (0,0) circle [radius=\R];
\draw[fill=white] (90:\R) circle [radius=\r];
\draw[fill=white] (-90:\R) circle [radius=\r];
\draw ([yshift=-\r cm]90:\R)--([yshift=\r cm]90:\R);
\draw ([yshift=-\r cm]-90:\R)--([yshift=\r cm]-90:\R);
\end{tikzpicture}}} \Bigg] 
\nonumber \\
&-\binom{n}{4}(n-5)! \vcenter{\hbox{
\begin{tikzpicture}
\def\R{.7};
\def\r{0.2};
\def\s{0.1cm};
\draw ([xshift=\s]-60:\R) arc (-60:60:\R);
\draw ([xshift=-\s]120:\R) arc (120:240:\R);
\draw[fill=white] ([xshift=\s]0:\R) circle [radius=\r];
\draw[fill=white] ([xshift=\s]60:\R) circle [radius=\r];
\draw[fill=white] ([xshift=-\s]120:\R) circle [radius=\r];
\draw[fill=white] ([xshift=-\s]180:\R) circle [radius=\r];
\draw[fill=white] ([xshift=-\s]240:\R) circle [radius=\r];
\draw[fill=white] ([xshift=\s]300:\R) circle [radius=\r];
\draw ([xshift=\s,yshift=-\r cm]0:\R)--([xshift=\s,yshift=\r cm]0:\R);
\draw ([xshift=\s,yshift=-\r cm]60:\R)--([xshift=\s,yshift=\r cm]60:\R);
\draw ([xshift=-\s,yshift=-\r cm]120:\R)--([xshift=-\s,yshift=\r cm]120:\R);
\draw ([xshift=-\s,yshift=-\r cm]180:\R)--([xshift=-\s,yshift=\r cm]180:\R);
\draw ([xshift=-\s,yshift=-\r cm]240:\R)--([xshift=-\s,yshift=\r cm]240:\R);
\draw ([xshift=\s,yshift=-\r cm]300:\R)--([xshift=\s,yshift=\r cm]300:\R);
\node at (270:\R) {\tiny{\dots}};
\node at (90:\R) {\tiny{\dots}};
\draw [decorate,decoration={brace,amplitude=8pt,mirror,raise=5pt}]
(-\R-\r,-\R) -- (\R+\r,-\R) node [black,midway,yshift=-0.8cm] 
{\footnotesize $(n-5) \ \text{terms}$}; 
\end{tikzpicture}}} \times \Bigg[ 3!
\vcenter{\hbox{
\begin{tikzpicture}
\def\R{.5};
\def\r{0.2};
\draw (0,0) circle [radius=\R];
\draw[fill=white] (45:\R) circle [radius=\r];
\draw[fill=white] (135:\R) circle [radius=\r];
\draw[fill=white] (225:\R) circle [radius=\r];
\draw[fill=white] (315:\R) circle [radius=\r];
\draw ([yshift=-\r cm]45:\R)--([yshift=\r cm]45:\R);
\draw ([yshift=-\r cm]135:\R)--([yshift=\r cm]135:\R);
\draw ([yshift=-\r cm]225:\R)--([yshift=\r cm]225:\R);
\draw ([yshift=-\r cm]315:\R)--([yshift=\r cm]315:\R);
\end{tikzpicture}}} -\frac{\binom{4}{2}\binom{2}{2}}{2!}
\vcenter{\hbox{
\begin{tikzpicture}
\def\R{.4};
\def\r{0.2};
\draw (0,0) circle [radius=\R];
\draw[fill=white] (90:\R) circle [radius=\r];
\draw[fill=white] (-90:\R) circle [radius=\r];
\draw ([yshift=-\r cm]90:\R)--([yshift=\r cm]90:\R);
\draw ([yshift=-\r cm]-90:\R)--([yshift=\r cm]-90:\R);
\end{tikzpicture}}}
\vcenter{\hbox{
\begin{tikzpicture}
\def\R{.4};
\def\r{0.2};
\draw (0,0) circle [radius=\R];
\draw[fill=white] (90:\R) circle [radius=\r];
\draw[fill=white] (-90:\R) circle [radius=\r];
\draw ([yshift=-\r cm]90:\R)--([yshift=\r cm]90:\R);
\draw ([yshift=-\r cm]-90:\R)--([yshift=\r cm]-90:\R);
\end{tikzpicture}}} \Bigg] 
\nonumber \\
&-\ldots-\binom{n}{n-2}
\vcenter{\hbox{
\begin{tikzpicture}
\def\R{.4};
\def\r{0.2};
\draw (0,0) circle [radius=\R];
\draw[fill=white] (90:\R) circle [radius=\r];
\draw[fill=white] (-90:\R) circle [radius=\r];
\draw ([yshift=-\r cm]90:\R)--([yshift=\r cm]90:\R);
\draw ([yshift=-\r cm]-90:\R)--([yshift=\r cm]-90:\R);
\end{tikzpicture}}} \times \Bigg[
(-1)^{\frac{n-2}{n}-1} \frac{\binom{n-2}{n}\binom{n-4}{n}\cdot\ldots\cdot\binom{2}{2}}{(\frac{n}{2})!} 
\vcenter{\hbox{
\begin{tikzpicture}
\def\R{.4cm};
\def\r{0.2};
\def\s{1cm};
\draw (0,0) circle [radius=\R];
\draw[fill=white] (90:\R) circle [radius=\r];
\draw[fill=white] (-90:\R) circle [radius=\r];
\draw ([yshift=-\r cm]90:\R)--([yshift=\r cm]90:\R);
\draw ([yshift=-\r cm]-90:\R)--([yshift=\r cm]-90:\R);
\draw ([xshift=\s]0,0) circle [radius=\R];
\draw[fill=white] ([xshift=\s]90:\R) circle [radius=\r];
\draw[fill=white] ([xshift=\s]-90:\R) circle [radius=\r];
\draw ([xshift=\s,yshift=-\r cm]90:\R)--([xshift=\s,yshift=\r cm]90:\R);
\draw ([xshift=\s,yshift=-\r cm]-90:\R)--([xshift=\s,yshift=\r cm]-90:\R);
\node at (1.5*\s+0.75*\R,0) {\ldots};
\draw ([xshift=2*\s+1.5*\R]0,0) circle [radius=\R];
\draw[fill=white] ([xshift=2*\s+1.5*\R]90:\R) circle [radius=\r];
\draw[fill=white] ([xshift=2*\s+1.5*\R]-90:\R) circle [radius=\r];
\draw ([xshift=2*\s+1.5*\R,yshift=-\r cm]90:\R)--([xshift=2*\s+1.5*\R,yshift=\r cm]90:\R);
\draw ([xshift=2*\s+1.5*\R,yshift=-\r cm]-90:\R)--([xshift=2*\s+1.5*\R,yshift=\r cm]-90:\R);
\draw [decorate,decoration={brace,amplitude=8pt,mirror,raise=8pt}]
(-\R,-\R) -- (2*\s+2.5*\R,-\R) node [black,midway,yshift=-0.8cm] 
{\footnotesize $(n-2)/2 \ \text{loops}$}; 
\end{tikzpicture}}} \ \Bigg] \Bigg\}. \label{detM}
\end{align}

\section{Explicit Expressions for Effective $SU(2)$ and $SU(3)$ Lagrangians}
In this section, we show the explicit calculation to obtain the effective Lagrangians for 2D topological field theory for $SU(2)$ and $SU(3)$ using the expression (\ref{effective}) together with the diagrammatic representation for adjugate matrix and matrix determinant expressed in (\ref{adjugate}) and (\ref{detM}) respectively.

For $SU(2)$, the adjugate matrix is
\begin{align}
    \text{adj}(\widetilde{M})\indices{^\alpha_\beta}= (-1) \raisebox{-0.1 cm}{\hbox{
\begin{tikzpicture}
\def\r{0.25};
\draw (0,0) circle [radius=\r];
\draw ([yshift=-\r cm]0,0)--([yshift=\r cm]0,0);
\node[above left] at (-0.7*\r,0) {\footnotesize{$\alpha$}};
\node[above right] at (0.7*\r,0) {\footnotesize{$\beta$}};
\end{tikzpicture}}}= -\hat{E}^{\alpha}_A \epsilon^{ACB}\phi\indices{_C} \hat{E}_{\beta B}
\end{align}
where $f^{ABC}=\epsilon^{ABC}$ for SU(2). Therefore, $\Theta^{AB}=-\epsilon^{ACB}\phi_C$. The matrix determinant is
\begin{align}
    \text{det}(\widetilde{M}) &= (-1) \vcenter{\hbox{
\begin{tikzpicture}
\def\R{.4};
\def\r{0.2};
\draw (0,0) circle [radius=\R];
\draw[fill=white] (90:\R) circle [radius=\r];
\draw[fill=white] (-90:\R) circle [radius=\r];
\draw ([yshift=-\r cm]90:\R)--([yshift=\r cm]90:\R);
\draw ([yshift=-\r cm]-90:\R)--([yshift=\r cm]-90:\R);
\end{tikzpicture}}}
= -\epsilon^{ABC}\phi\indices{_B}\eta\indices{_C_D}\epsilon^{DEF}\phi\indices{_E}\eta\indices{_A_F} \nonumber \\
&= 2\eta^{BD}\phi\indices{_B}\phi\indices{_D} =2 |\phi|^2.
\end{align}
Thus, when substituting the above relations into (\ref{effective}), we obtain
\begin{equation}
\mathcal{L}_\text{eff}(\phi)= \  \frac{-i}{g|\phi|^2}\partial\phi\indices{_A} \epsilon^{ABC}\phi\indices{_B} \bar{\partial}\phi\indices{_C}
\end{equation}
which is identical to what we found earlier in the equation (\ref{SU2 eff}).

For $SU(3)$, the diagrammatic expressions for the adjugate matrix and matrix determinant are 
\begin{align}
\text{adj}(\widetilde{M})\indices{^\alpha_\beta}=&(-1)\Bigg\{5! \raisebox{-0.1 cm}{\hbox{\begin{tikzpicture}
\def\r{0.25};
\def\l{0.4};
\draw (0,0) circle [radius=\r];
\draw (0,-\r)--(0,\r);
\draw (\r,0)--(\r+\l,0) (-\r,0)--(-\r-\l,0);
\draw (-\l-2*\r,0) circle [radius=\r];
\draw (-\l-2*\r,-\r)--(-\l-2*\r,\r);
\node[above left] at (-\l-2.7*\r,0) {\footnotesize{$\alpha$}};
\draw (2*\r+\l,0) circle [radius=\r];
\draw (2*\r+\l,-\r)--(2*\r+\l,\r);
\draw (3*\r+\l,0)--(3*\r+2*\l,0);
\draw (4*\r+2*\l,0) circle [radius=\r];
\draw (4*\r+2*\l,-\r)--(4*\r+2*\l,\r);
\draw (5*\r+2*\l,0)--(5*\r+3*\l,0);
\draw (6*\r+3*\l,0) circle [radius=\r];
\draw (6*\r+3*\l,-\r)--(6*\r+3*\l,\r);
\node[above right] at (3*\l+6.7*\r,0) {\footnotesize{$\beta$}};
\end{tikzpicture}}} \nonumber \\
&-3! \raisebox{-0.1 cm}{\hbox{\begin{tikzpicture}
\def\r{0.25};
\def\l{0.4};
\draw (0,0) circle [radius=\r];
\draw (0,-\r)--(0,\r);
\draw (\r,0)--(\r+\l,0) (-\r,0)--(-\r-\l,0);
\draw (-\l-2*\r,0) circle [radius=\r];
\draw (-\l-2*\r,-\r)--(-\l-2*\r,\r);
\node[above left] at (-\l-2.7*\r,0) {\footnotesize{$\alpha$}};
\draw (2*\r+\l,0) circle [radius=\r];
\draw (2*\r+\l,-\r)--(2*\r+\l,\r);
\node[above right] at (1*\l+2.7*\r,0) {\footnotesize{$\beta$}};
\end{tikzpicture}}} \times \binom{5}{2}
\vcenter{\hbox{
\begin{tikzpicture}
\def\R{.4};
\def\r{0.2};
\draw (0,0) circle [radius=\R];
\draw[fill=white] (90:\R) circle [radius=\r];
\draw[fill=white] (-90:\R) circle [radius=\r];
\draw ([yshift=-\r cm]90:\R)--([yshift=\r cm]90:\R);
\draw ([yshift=-\r cm]-90:\R)--([yshift=\r cm]-90:\R);
\end{tikzpicture}}} \nonumber \\
&-1! \raisebox{-0.1 cm}{\hbox{
\begin{tikzpicture}
\def\r{0.25};
\draw (0,0) circle [radius=\r];
\draw ([yshift=-\r cm]0,0)--([yshift=\r cm]0,0);
\node[above left] at (-0.7*\r,0) {\footnotesize{$\alpha$}};
\node[above right] at (0.7*\r,0) {\footnotesize{$\beta$}};
\end{tikzpicture}}} \times \Bigg[ 3!
\vcenter{\hbox{
\begin{tikzpicture}
\def\R{.5};
\def\r{0.2};
\draw (0,0) circle [radius=\R];
\draw[fill=white] (45:\R) circle [radius=\r];
\draw[fill=white] (135:\R) circle [radius=\r];
\draw[fill=white] (225:\R) circle [radius=\r];
\draw[fill=white] (315:\R) circle [radius=\r];
\draw ([yshift=-\r cm]45:\R)--([yshift=\r cm]45:\R);
\draw ([yshift=-\r cm]135:\R)--([yshift=\r cm]135:\R);
\draw ([yshift=-\r cm]225:\R)--([yshift=\r cm]225:\R);
\draw ([yshift=-\r cm]315:\R)--([yshift=\r cm]315:\R);
\end{tikzpicture}}} -\frac{\binom{4}{2}\binom{2}{2}}{2!}
\vcenter{\hbox{
\begin{tikzpicture}
\def\R{.4};
\def\r{0.2};
\draw (0,0) circle [radius=\R];
\draw[fill=white] (90:\R) circle [radius=\r];
\draw[fill=white] (-90:\R) circle [radius=\r];
\draw ([yshift=-\r cm]90:\R)--([yshift=\r cm]90:\R);
\draw ([yshift=-\r cm]-90:\R)--([yshift=\r cm]-90:\R);
\end{tikzpicture}}}
\vcenter{\hbox{
\begin{tikzpicture}
\def\R{.4};
\def\r{0.2};
\draw (0,0) circle [radius=\R];
\draw[fill=white] (90:\R) circle [radius=\r];
\draw[fill=white] (-90:\R) circle [radius=\r];
\draw ([yshift=-\r cm]90:\R)--([yshift=\r cm]90:\R);
\draw ([yshift=-\r cm]-90:\R)--([yshift=\r cm]-90:\R);
\end{tikzpicture}}} \Bigg] \Bigg\}
\end{align}
and
\begin{align}
\det(\widetilde{M})=(-1)\Bigg\{ &5! \vcenter{\hbox{
\begin{tikzpicture}
\def\R{.7};
\def\r{0.2};
\def\s{0cm};
\draw (0,0) circle [radius=\R];
\draw[fill=white] (0:\R) circle [radius=\r];
\draw[fill=white] (60:\R) circle [radius=\r];
\draw[fill=white] (120:\R) circle [radius=\r];
\draw[fill=white] (180:\R) circle [radius=\r];
\draw[fill=white] (240:\R) circle [radius=\r];
\draw[fill=white] (300:\R) circle [radius=\r];
\draw ([yshift=-\r cm]0:\R)--([yshift=\r cm]0:\R);
\draw ([yshift=-\r cm]60:\R)--([yshift=\r cm]60:\R);
\draw ([yshift=-\r cm]120:\R)--([yshift=\r cm]120:\R);
\draw ([yshift=-\r cm]180:\R)--([yshift=\r cm]180:\R);
\draw ([yshift=-\r cm]240:\R)--([yshift=\r cm]240:\R);
\draw ([yshift=-\r cm]300:\R)--([yshift=\r cm]300:\R);
\end{tikzpicture}}} 
-3!\vcenter{\hbox{
\begin{tikzpicture}
\def\R{.5};
\def\r{0.2};
\draw (0,0) circle [radius=\R];
\draw[fill=white] (45:\R) circle [radius=\r];
\draw[fill=white] (135:\R) circle [radius=\r];
\draw[fill=white] (225:\R) circle [radius=\r];
\draw[fill=white] (315:\R) circle [radius=\r];
\draw ([yshift=-\r cm]45:\R)--([yshift=\r cm]45:\R);
\draw ([yshift=-\r cm]135:\R)--([yshift=\r cm]135:\R);
\draw ([yshift=-\r cm]225:\R)--([yshift=\r cm]225:\R);
\draw ([yshift=-\r cm]315:\R)--([yshift=\r cm]315:\R);
\end{tikzpicture}}} \times \binom{6}{2}
\vcenter{\hbox{
\begin{tikzpicture}
\def\R{.4};
\def\r{0.2};
\draw (0,0) circle [radius=\R];
\draw[fill=white] (90:\R) circle [radius=\r];
\draw[fill=white] (-90:\R) circle [radius=\r];
\draw ([yshift=-\r cm]90:\R)--([yshift=\r cm]90:\R);
\draw ([yshift=-\r cm]-90:\R)--([yshift=\r cm]-90:\R);
\end{tikzpicture}}} 
\nonumber \\
&-\frac{\binom{6}{2}\binom{4}{2}}{3!}
\vcenter{\hbox{
\begin{tikzpicture}
\def\R{.4};
\def\r{0.2};
\draw (0,0) circle [radius=\R];
\draw[fill=white] (90:\R) circle [radius=\r];
\draw[fill=white] (-90:\R) circle [radius=\r];
\draw ([yshift=-\r cm]90:\R)--([yshift=\r cm]90:\R);
\draw ([yshift=-\r cm]-90:\R)--([yshift=\r cm]-90:\R);
\end{tikzpicture}}}
\vcenter{\hbox{
\begin{tikzpicture}
\def\R{.4};
\def\r{0.2};
\draw (0,0) circle [radius=\R];
\draw[fill=white] (90:\R) circle [radius=\r];
\draw[fill=white] (-90:\R) circle [radius=\r];
\draw ([yshift=-\r cm]90:\R)--([yshift=\r cm]90:\R);
\draw ([yshift=-\r cm]-90:\R)--([yshift=\r cm]-90:\R);
\end{tikzpicture}}}
\vcenter{\hbox{
\begin{tikzpicture}
\def\R{.4};
\def\r{0.2};
\draw (0,0) circle [radius=\R];
\draw[fill=white] (90:\R) circle [radius=\r];
\draw[fill=white] (-90:\R) circle [radius=\r];
\draw ([yshift=-\r cm]90:\R)--([yshift=\r cm]90:\R);
\draw ([yshift=-\r cm]-90:\R)--([yshift=\r cm]-90:\R);
\end{tikzpicture}}} \Bigg\}
\end{align}

According to the above expressions, one can write the effective Lagrangian in the form (\ref{effective}) with 
\begin{align}
    \Theta^{AB}=&- 5! (\mathcal{F}\indices{^A_C}\mathcal{F}\indices{^C_D}\mathcal{F}\indices{^D_E}\mathcal{F}\indices{^E_F}\mathcal{F}\indices{^F^B})+3!\cdot10 \cdot (\mathcal{F}\indices{^A_C}\mathcal{F}\indices{^C_D}\mathcal{F}\indices{^D^B})(\mathcal{F}\indices{^E_F}\mathcal{F}\indices{^F_E}) \nonumber \\
    &+\mathcal{F}\indices{^A^B}\Big[ 3! (\mathcal{F}\indices{^C_D}\mathcal{F}\indices{^D_E}\mathcal{F}\indices{^E_F}\mathcal{F}\indices{^E_C})-3(\mathcal{F}\indices{^C_D}\mathcal{F}\indices{^D_C})(\mathcal{F}\indices{^E_F}\mathcal{F}\indices{^F_E}) \Big] \label{su3 adj}
\end{align}
and
\begin{align}
    \det(\widetilde{M})=&- 5! (\mathcal{F}\indices{^A_B}\mathcal{F}\indices{^B_C}\mathcal{F}\indices{^C_D}\mathcal{F}\indices{^E_F}\mathcal{F}\indices{^F_G}\mathcal{F}\indices{^G_A})+3!\cdot15 \cdot (\mathcal{F}\indices{^A_B}\mathcal{F}\indices{^B_C}\mathcal{F}\indices{^C_D}\mathcal{F}\indices{^D_A})(\mathcal{F}\indices{^E_F}\mathcal{F}\indices{^F_E}) \nonumber \\
    &+15 (\mathcal{F}\indices{^A_B}\mathcal{F}\indices{^B_A})(\mathcal{F}\indices{^C_D}\mathcal{F}\indices{^D_C})(\mathcal{F}\indices{^E_F}\mathcal{F}\indices{^F_E}) \label{su3 det}
\end{align}
where we used the notation $\mathcal{F}^{AB}=f^{ACB}\phi\indices{_C}$.

We can further simplify the above terms by expanding them explicitly in the Cartan-Weyl basis for $SU(3)$. The generators are

\begin{align}
& I^+= \begin{pmatrix}
0&1&0 \\ 0&0&0 \\ 0&0&0
\end{pmatrix},
&I^-= \begin{pmatrix}
0&0&0 \\ 1&0&0 \\ 0&0&0 \end{pmatrix}, & & I^3=\frac{1}{2} \begin{pmatrix}
1&0&0 \\ 0&-1&0 \\ 0&0&0
\end{pmatrix},
&& \nonumber \\
&U^+= \begin{pmatrix}
0&0&0 \\ 0&0&1 \\ 0&0&0
\end{pmatrix},
&U^-= \begin{pmatrix}
0&0&0 \\ 0&0&0 \\ 0&1&0 \end{pmatrix}, & &
&& \nonumber \\
&V^+=\frac{1}{2} \begin{pmatrix}
0&0&1 \\ 0&0&0 \\ 0&0&0 \end{pmatrix},
&
V^-= \begin{pmatrix}
0&0&0 \\ 0&0&0 \\ 1&0&0
\end{pmatrix}, & &Y=\frac{1}{3} \begin{pmatrix}
1&0&0 \\ 0&1&0 \\ 0&0&-2 \end{pmatrix}.
&&
\end{align}
We can determine the structure constants by considering all matrix commutators between the elements. The generators $I^3$ and $Y$ are the Cartan subalgebra elements satisfying 
\begin{equation}
[I^3,Y]=0.
\end{equation}
The plus and minus superscripts of the generators denote the raising and lowering operators within the three su(2) subalgebras given by
\begin{align}
    [I^+,I^-]=2I^3, \qquad [U^+,U^-]=\frac{3}{2}Y-I^3, \qquad  [V^+,V^-]=\frac{3}{2}Y+I^3. \label{subalgebra}
\end{align}
Note that the Hermitian conjugation of generators switches the plus and minus superscripts of the generators within each SU(2) subgroup, i.e.  $(I^\pm)^\dag=I^\mp$, $(U^\pm)^\dag=U^\mp$, $(V^\pm)^\dag=V^\mp$.

Apart from (\ref{subalgebra}), the remaining non-zero commutators are
\begin{align}
    [I^3,I^\pm]&=\pm I^\pm, &[I^3,U^\pm]&=\mp\frac{1}{2}U^\pm, \nonumber \\
    [I^3,V^\pm]&=\pm\frac{1}{2}U^\pm, &[Y,U^\pm]&=\pm U^\pm, \nonumber \\
    [Y,V^\pm]&=\pm V^\pm, &[I^\pm,U^\pm]&=\pm V^\pm, \nonumber \\
    [I^\pm,V^\mp]&=\mp U^\mp, &[U^\pm,V^\mp]&=\pm I^\mp.
\end{align}

For convenience, we denote  $\{I^3,Y,I^+,I^-,U^+,U^-,V^+,V^-  \}$ by $\{T^1,T^2, \ldots ,T^8  \}$ respectively. In this notation, the metric tensor $\eta^{AB}$ can be written as 
\begin{equation}
\setlength\arraycolsep{5pt}
    \eta^{AB}=\begin{pmatrix}
    1 & 0 & 0 & 0 & 0 & 0 & 0 & 0 \\
    0 & \frac{4}{3} & 0 & 0 & 0 & 0 & 0 & 0 \\
    0 & 0 & 0 & 2 & 0 & 0 & 0 & 0 \\
    0 & 0 & 2 & 0 & 0 & 0 & 0 & 0 \\
    0 & 0 & 0 & 0 & 0 & 2 & 0 & 0 \\
    0 & 0 & 0 & 0 & 2 & 0 & 0 & 0 \\
    0 & 0 & 0 & 0 & 0 & 0 & 0 & 2 \\
    0 & 0 & 0 & 0 & 0 & 0 & 2 & 0 \\                    
    \end{pmatrix}
\end{equation}
which is directly from $\eta^{AB}=2 \ \text{tr}(T^AT^B)$.

The field $\phi$ is an element of the Cartan subalgebra, i.e. $\phi=\phi_1 T^1+\phi_2 T^2$. Consequently, one can find the adjoint representation of the field $\phi$ as

\begin{align}
    &\text{ad}(\phi)= \nonumber \\
    &\begin{pmatrix}
    \phantom{00}0\phantom{00}&  \phantom{00}0\phantom{00}&  \phantom{00}0\phantom{00}&  \phantom{00}0\phantom{00}&  0&  0&  0&  0 \\
    0&  0&  0&  0&  0&  0&  0&  0 \\
    0&  0&  0&  -2\phi\indices{_1}&  0&  0&  0&  0 \\
    0&  0&  2\phi\indices{_1}&  0&  0&  0&  0&  0 \\
    0&  0&  0&  0&   0&  \phi\indices{_1}-2\phi\indices{_2}&  0&  0 \\
    0&  0&  0&  0&  -\phi\indices{_1}+2\phi\indices{_2}&  0&  0&  0 \\
    0&  0&  0&  0&  0&  0&  0&  -\phi\indices{_1}-2\phi\indices{_2} \\
    0&  0&  0&  0&  0&  0&  \phi\indices{_1}+2\phi\indices{_2}&  0 \\
    \end{pmatrix} \label{adjoint}
\end{align}
where $\text{ad}(\phi)=if^{ABC}\phi\indices{_B}=i\mathcal{F}^{AC}$. With this matrix (\ref{adjoint}), one can compute all loop and strand diagrams appearing in the equations (\ref{su3 adj}) and (\ref{su3 det}). Chains of matrix multiplications of the matrix $\mathcal{F}$ are shown in the Appendix.

From the calculation, we can further simplify the loop terms. The loop diagram with two vertices $\mathcal{F}\indices{^A_B}\mathcal{F}\indices{^B_C}$ can be replaced by the absolute square of the field $\phi$ as
\begin{align}
        \mathcal{F}\indices{^A_B}\mathcal{F}\indices{^B_A}&=-2(\phi\indices{_1})^2-\frac{1}{2}(\phi\indices{_1}-2\phi\indices{_2})^2-\frac{1}{2}(\phi\indices{_1}+2\phi\indices{_2})^2 \nonumber \\
        &=-3((\phi\indices{_1})^2+\frac{4}{3}(\phi\indices{_2})^2) \ = \ -3|\phi|^2.
\end{align}
A similar pattern appears in the four-vertex loop as it is proportional to $|\phi|^4$:
\begin{align}
        \mathcal{F}\indices{^A_B}\mathcal{F}\indices{^B_C}\mathcal{F}\indices{^C_D}\mathcal{F}\indices{^D_A}&=2(\phi\indices{_1})^4+\frac{1}{8}(\phi\indices{_1}-2\phi\indices{_2})^4+\frac{1}{8}(\phi\indices{_1}+2\phi\indices{_2})^4 \nonumber \\
        &=\frac{9}{4}((\phi\indices{_1})^2+\frac{4}{3}(\phi\indices{_2})^2)^2 \ = \ \frac{9}{4}|\phi|^4.
\end{align}
The six-vertex loop can be expressed in terms of two invariant objects, $|\phi|^6$ and $d^{ABC}\phi\indices{_A}\phi\indices{_B}\phi\indices{_C}$ as
\begin{align}
        \mathcal{F}\indices{^A_B}\mathcal{F}\indices{^B_C}\mathcal{F}\indices{^C_D}\mathcal{F}\indices{^D_E}\mathcal{F}\indices{^E_F}\mathcal{F}\indices{^F_A}&=-2(\phi\indices{_1})^6-\frac{1}{32}(\phi\indices{_1}-2\phi\indices{_2})^6-\frac{1}{32}(\phi\indices{_1}+2\phi\indices{_2})^6 \nonumber \\
        &=-\frac{33}{16}|\phi|^6+\frac{9}{8}\bigg( 2(\phi\indices{_1})^2(\phi\indices{_3})-\frac{8}{9}(\phi\indices{_2})^3 \bigg)^2
\end{align}
where the quantity inside the parenthesis is $d^{ABC}\phi\indices{_A}\phi\indices{_B}\phi\indices{_C}$ where $d^{ABC}$ is the totally symmetric third rank tensor defined in (\ref{d factor}).

In consequence, we can rewrite the terms (\ref{su3 adj}) and (\ref{su3 det}) as
\begin{align}
    \Theta^{AB}=&- 120 (\mathcal{F}\indices{^A_C}\mathcal{F}\indices{^C_D}\mathcal{F}\indices{^D_E}\mathcal{F}\indices{^E_F}\mathcal{F}\indices{^F^B})-180 (\mathcal{F}\indices{^A_C}\mathcal{F}\indices{^C_D}\mathcal{F}\indices{^D^B})|\phi|^2
    -\frac{27}{2}\mathcal{F}\indices{^A^B}|\phi|^4 \label{su3 adj2}
\end{align}
and
\begin{align}
    \det(\widetilde{M})=&-765|\phi|^6-135 (d^{ABC}\phi\indices{_A}\phi\indices{_B}\phi\indices{_C}) \label{su3 det2}.
\end{align}
\section{The Topological Field Action with a Source Term and the Expectation Value of the Wilson Loop}
In this section we would like to generalise the action (\ref{Topological field action}) further by adding a source term for the gauge field $\mathcal{A}$. Consider the action given by
\begin{equation}
    S[\mathcal{J}]=2\int_\mathcal{M} d^2\xi \ \epsilon^{ij} \text{tr}(\phi \mathcal{F}_{ij}+\mathcal{J}_i\mathcal{A}_j). \label{action with source}
\end{equation}

To obtain the effective Lagrangian for the field $\phi$, we will integrate out the gauge field $\mathcal{A}$ as before. By doing so, we expand the field $\mathcal{A}$ in terms of the unit basis defined by (\ref{Lie comp}). This gives the partition function as
\begin{equation}
    Z[\mathcal{J}]=\frac{1}{\text{Vol}}\int D\phi\indices{_A} Da^\alpha_i  D\chi_{ja} \ \text{exp}(-\Tilde{S}[\mathcal{J}]). \label{partition function with source}
\end{equation}
with
\begin{align}
    \tilde{S}[\mathcal{J}]=\int_\mathcal{M} d^2\xi  \Big(&ig f^{ABC} \phi\indices{_C} a^\alpha_{i} a^\beta_{j} \hat{E}_{\alpha A}\hat{E}_{\beta B}  -(2\partial_i\phi_A-\mathcal{J}_{iA})a^\alpha_j\hat{E}^A_\alpha \nonumber \\
    &-(2\partial_i\phi_A-\mathcal{J}_{iA})\chi_{ja}\hat{H}^{Aa}\Big) \epsilon^{ij}. 
\end{align}
It is not hard to see that the path integration of the last line leads to a constraint on the theory. This appears in the form of a Dirac delta function 
\begin{equation}
    \prod_{a=1}^{N-1}\prod_{i=1}^2 \delta (\text{tr}(2\partial_i\phi-\mathcal{J}_i)H^a).
\end{equation}
The constraint implies that the difference between  $2\partial\phi$ and $\mathcal{J}$ does not lie in the Cartan subalgebra.

We can proceed with the calculation as in previous sections by changing from spacetime coordinates $(\xi^1,\xi^2)$ to the complex coordinates $(z,\bar{z})$. The partition function now resembles a Gaussian path integral with respect to the complex fields $b$ and $\bar{b}$ expressed in (\ref{complex b field}) which is
\begin{equation}
    Z[\mathcal{J}]=\frac{\mathcal{N}}{\text{Vol}}\int D\phi\indices{_A} Db^\alpha D\bar{b}^\alpha \ \prod_{i=1}^2\delta^{(N-1)}(\text{tr}((2\partial_i\phi-\mathcal{J}_i)\hat\phi))\text{exp}(-S[\phi,b,\bar{b},\mathcal{J}]). \label{partition function with souce complex}
\end{equation}
where
\begin{equation}
    S[\phi,b,\bar{b},\mathcal{J}]=\int_\mathcal{M} d^2z \Big( 2igf^{ABC}\phi\indices{_C} b^\alpha\bar{b}^\beta \hat{E}_{\alpha A}\hat{E}_{\beta B} -((2\partial\phi_A-\mathcal{J}_{A})\bar{b}^\alpha-(2\bar{\partial}\phi_A-\bar{\mathcal{J}}_A)b^\alpha)\hat{E}^A_\alpha \Big).
\end{equation}

Integrating out the $b$ and $\bar{b}$ using (\ref{gaussian}) yields
\begin{equation}
    Z[\mathcal{J}]=\frac{\mathcal{N}}{\text{Vol}}\int D\phi\indices{_A}  \ \prod_{i=1}^2\delta^{(N-1)}(\text{tr}((2\partial_i\phi-\mathcal{J}_i)\hat\phi))\text{exp}\Big(-S_\text{eff}(\phi,\mathcal{J})\Big) \label{partition function with souce complex2}
\end{equation}
with
\begin{equation}
S_\text{eff}(\phi,\mathcal{J})=\int d^2z \ \frac{i}{2g} \frac{1}{\text{det}(\widetilde{M})}(2\partial\phi\indices{_A}-\mathcal{J}_A) \Theta^{AB} (2\bar{\partial}\phi\indices{_B}-\bar{\mathcal{J}}_B).  \label{complex effective with source}
\end{equation}
Turning back to the $(\xi^1,\xi^2)$ coordinates, the effective action takes the form
\begin{equation}
S_\text{eff}(\phi,\mathcal{J})=\int d^2 \xi \ \frac{i}{4g} \frac{1}{\text{det}(\widetilde{M})}(2\partial_i\phi\indices{_A}-\mathcal{J}_{iA}) \Theta^{AB} (2\partial_j\phi\indices{_B}-\mathcal{J}_{jB})\epsilon^{ij}.  \label{effective with source}
\end{equation}

It is known that one can relate a BF theory to 2D Yang-Mills theory by introducing the quadratic term for the field $\phi$ which is 
\begin{equation}
    S_{qd}=e^2\int d^2 \xi \sqrt{g}|\phi|^2. \label{mass term}
\end{equation}
As a result, the partition function for the 2D gauge theory with the gauge field source $\mathcal{J}$ can be expressed as
\begin{equation}
        Z[\mathcal{J}]=\frac{\mathcal{N}}{\text{Vol}}\int D\phi\indices{_A}  \ \prod_{i=1}^2\delta^{(N-1)}(\text{tr}((2\partial_i\phi-\mathcal{J}_i)\hat\phi))\text{exp}\Big(-(S_\text{eff}+S_{qd}) \Big). \label{partition function gauge theory}
\end{equation}

With a suitable choice of the source term $\mathcal{J}$, in principle we are able to compute the expectation value of a Wilson loop in 2D Yang-Mills theory based on our effective BF theory. However, we have to deal with the issue of path-ordering.

The non-Abelian Wilson loop can be expressed as the trace of the path-ordered exponential of a line integral of the gauge field $\mathcal{A}$ along a closed loop $C$,
\begin{equation}
    W[C]=\text{tr}\Big( \mathcal{P}\Big(e^{-g\oint_C \mathcal{A}\cdot d\xi} \Big) \Big).
\end{equation}
The trace together with the path-ordering operator can be replaced by a functional integral over a complex anti-commuting field $\psi$ \cite{Samuel:1978iy,Broda:1995wv} as
\begin{equation}
    W[C]= \int D\psi^\dagger D\psi \exp \Big(\int d\tau \psi^\dagger \dot{\psi} -g\mathcal{A}_{iR} \dot{\xi}^i \psi^\dagger T^R \psi \Big)
\end{equation}
where the loop $C$ is now parameterized by $\tau$. Therefore, the expectation value of the Wilson loop takes the form 
\begin{align}
    \langle W[C] \rangle=\frac{1}{Z'}\int D\phi D\psi^\dagger D\psi \prod_{i=1}^2&\delta^{(N-1)}(\text{tr}((2\partial_i\phi-\mathcal{J}_i)\hat\phi)) \nonumber \\  &\times\exp{\big({-(S_\text{eff}+S_{qd})+\int d\tau \psi^\dagger \dot{\psi}}\big)} \label{exp value Wilson}
\end{align}
with
\begin{equation}
    \mathcal{J}\indices{^A_i}(\xi)=-g\oint_{C} \psi^\dagger(\tilde{\xi}) T^A \psi(\tilde{\xi}) \delta^{(2)}(\xi-\tilde{\xi})\epsilon_{ij} d\tilde{\xi}^j \label{wilson loop source}
\end{equation}
and the action $S_\text{eff}$ and $S_{qd}$ are expressed in (\ref{effective with source}) and (\ref{mass term}) respectively. The term $Z'$ in the denominator is a normalization factor such that $\langle 1 \rangle=1$.

However, it turns out that the solution for the equation $2\partial_i\phi-\mathcal{J}_i=0$ with the source term expressed above is not consistent as the line integral of $\mathcal{J}_i$ is not path independent which contradicts to the equation itself. To deal with this, we will exploit gauge symmetry.

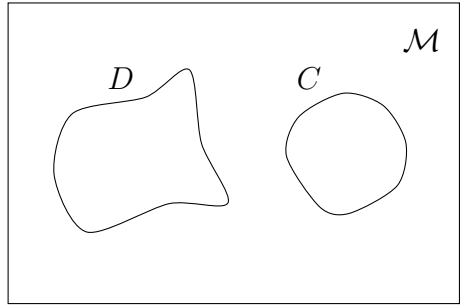
\begin{figure}[t]
    \centering
\begin{tikzpicture}
\draw (-3,-2)--(3,-2)--(3,2)--(-3,2)--cycle;
  \pgfmathsetseed{1}
\draw plot [smooth cycle, samples=8,domain={1:8},xshift=-1cm] (\x*360/8+16*rnd:0.5cm+1cm*rnd) node at (-1.5,1) {$D$};
\draw plot [smooth cycle, samples=8,domain={1:8},xshift=1.5cm] (\x*360/8+20*rnd:0.8cm);
\node at (1,1) {$C$};
\node at (2.5,1.5) {$\mathcal{M}$};
\end{tikzpicture}
    \caption{Two-dimensional manifold $\mathcal{M}$ with a region $D$ and a closed loop $C$}
    \label{manifold}
\end{figure}

For simplicity, we will proceed with the calculation in the context of $SU(2)$ theory. In this setting, we choose the gauge fixing such that the unit vector $\hat\phi$ is constant everywhere outside a region $D$. Therefore, the manifold $\mathcal{M}$ now consists of the region $D$ where the value of $\hat\phi$ varies and the rest of the manifold where the $\hat\phi$ is constant. We can further choose that the region $D$ does not intercept the loop $C$ as depicted in the figure \ref{manifold}.

According to this gauge choice, the effective action term (\ref{effective with source}) becomes
\begin{align}
    S_\text{eff}(\phi,\mathcal{J})=&\int_D d^2 \xi \ \frac{i|\phi|}{2g} \partial_i\hat\phi\indices{_A}  \partial_j\hat\phi\indices{_B}\hat\phi\indices{_C}\epsilon^{ABC}\epsilon^{ij} \nonumber\\
   & +\int_{\mathcal{M}/D} d^2 \xi \ \frac{i}{8g|\phi|^2} (2\partial_i\phi\indices{_A}-\mathcal{J}_{iA})  (2\partial_j\phi\indices{_B}-\mathcal{J}_{jB})\phi\indices{_C}\epsilon^{ABC}\epsilon^{ij}. \label{effective gauged action}
\end{align}
If we consider the case when the manifold $\mathcal{M}$ has a topology of unit sphere $S^2$, the first term can be related to a winding number as discussed in the earlier section. Note that $|\phi|$ is constant due to the absence of the source in $D$. Moreover, since the $\hat\phi$ is constant in $\mathcal{M}/D$, the non-vanishing contribution to the second line is
\begin{align}
        \int_{\mathcal{M}/D}& d^2 \xi \ \frac{i}{8g|\phi|^2} \mathcal{J}_{iA}  \mathcal{J}_{jB}\phi\indices{_C}\epsilon^{ABC}\epsilon^{ij} \nonumber \\
        &=\frac{ig}{8} \oint_C \oint_C d\tilde{\xi}^i d\xi'^j \bigg( \psi^\dagger T^A \psi \biggr\rvert_{\tilde{\xi}}\bigg) \bigg( \psi^\dagger T^B \psi \biggr\rvert_{\xi'} \bigg) \delta^{(2)}(\tilde{\xi}-\xi') \epsilon_{ij} \frac{\phi^C}{|\phi|^2}  \epsilon\indices{_A_B_C}.
\end{align}
The term $\oint_C\oint_C d\tilde{\xi}^i d\xi'^j \delta^{(2)}(\tilde{\xi}-\xi') \epsilon_{ij}$ counts the number of times the loop $C$ intersects itself. Therefore, the above term can be set to zero provided that the loop $C$ does not have a self-intersection. Subsequently, the effective action (\ref{effective gauged action}) turns into
\begin{equation}
    S_\text{eff}(\phi)=\frac{i}{g}|\phi|(4\pi n) \label{winding number action}
\end{equation}
where $n$ is the winding number of the map $\hat\phi$.

At this point, the appearance of the fermionic field $\psi$ in the effective action $S_{\text{eff}}$ has been removed due to the gauge choice. Therefore, according to (\ref{exp value Wilson}), the only term that is subject to the path-ordering operation is the source term $\mathcal{J}$ in the constraint. This allows us to rewrite (\ref{exp value Wilson}) as
\begin{align}
    \langle W[C] \rangle=\frac{1}{Z}\int  D\phi D\chi \ \text{tr}\bigg[\mathcal{P}\bigg(  \exp\bigg({\frac{g}{2}\oint_C \hat{\phi}\chi_i d\xi^i} \bigg)\bigg)\bigg] \nonumber \\ 
    \times \exp{\bigg(-(S_\text{eff}+S_{qd})+\int d^2\xi (\partial_i \phi\indices{_A})\hat\phi\indices{^A}\chi_j\epsilon^{ij}} \bigg)
\end{align}
where the Dirac delta function is replaced by the functional integral over the field $\chi$. Since the field $\hat\phi$ is constant and commutes with itself throughout the loop, the path-ordering operator $\mathcal{P}$ can be dropped. Denoting the eigenvalue of $\hat\phi$ by $\lambda$, the trace of the exponential in the first line takes the form
\begin{equation}
    \sum_\lambda \exp\bigg( \frac{g\lambda}{2}\int d^2\xi\oint_C \delta^{(2)}(\xi-\tilde{\xi})\chi_i(\xi)d\tilde{\xi}^i \bigg).
\end{equation}

We then proceed with the calculation by integrating out the field $\chi$. This generates a constraint via a Dirac delta function as
\begin{align}
    \langle W[C] \rangle=\frac{1}{Z}\int  D\phi  \ \sum_\lambda \prod_{i=1}^2\delta\bigg(\partial_i|\phi|+\frac{g\lambda}{2}\oint_C \delta^{(2)}(\xi-\tilde{\xi})  \epsilon_{ij}d\tilde{\xi}^j \bigg) e^{-(S_\text{eff}+S_{qd})}. \label{wilson loop expectation}
\end{align}

It is not hard to see that the solution for the constraint,
\begin{equation}
    \partial_i|\phi|+\frac{g\lambda}{2}\oint_C \delta^{(2)}(\xi-\tilde{\xi})  \epsilon_{ij}d\tilde{\xi}^j=0. \label{constraint wilson loop}
\end{equation}
takes the form
\begin{equation}
    \varphi_\lambda-\varphi_0=-\frac{g\lambda}{2} \bigg( \int_O^\xi \oint_{C} \delta^{(2)}(\xi'-\tilde{\xi})\epsilon_{ij} d\tilde{\xi}^i d\xi'^j \bigg).
\end{equation}
where $\varphi_\lambda$ and $\varphi_0$ are the scalar fields at arbitrary point $\xi$ and a reference point $O$ respectively. The object in the parenthesis counts the number of oriented intersections between two curves \cite{Curry:2017cnu}. The solution above is independent of path, hence, it depends only on the reference point $O$. If we set the point $O$ to be outside the loop $C$,
\begin{equation}
  \varphi_\lambda-\varphi_0=\begin{cases}
    -\frac{g\lambda}{2}, & \text{if $\xi$ is inside the loop $C$}\\
    0, & \text{otherwise}.
  \end{cases}\label{varphi}
\end{equation}

This allow us to compute the expectation value of the Wilson loop in 2D Yang-Mills theory (\ref{wilson loop expectation}) as

\begin{align}
    \langle W[C] \rangle&=\frac{1}{Z}\sum_\lambda \int_{0}^\infty d\varphi_0 \sum_{n=-\infty}^\infty \exp\bigg[{\frac{-i}{g}(4\pi n)\varphi_0-e^2 \int_\mathcal{M} d^2\xi \sqrt{g}\varphi^2_\lambda}\bigg]  \label{avg Wilson loop}
\end{align}
The infinite $m$ limit of the Dirichlet kernel, $D_m(x)$, represents the Dirac delta function as
\begin{equation}
    \lim_{m\rightarrow \infty} D_m(x)=\lim_{m\rightarrow \infty} \sum_{k=-m}^m e^{imx}=2\pi \delta(x)
\end{equation}
where $x\in [0,2\pi]$. Therefore, (\ref{avg Wilson loop}) becomes
\begin{align}
    \langle W[C] \rangle=&\frac{1}{Z}\sum_\lambda \int_{0}^\infty d\varphi_0 \ \frac{g}{2}\delta(\varphi_0 \bmod \frac{g}{2}) \nonumber \\
    &\times\exp\bigg[-e^2 \bigg( \int_\Gamma d^2\xi \sqrt{g}\varphi^2_\lambda+ \int_{\mathcal{M}/\Gamma} d^2\xi \sqrt{g}\varphi^2_\lambda\bigg)\bigg]  \label{avg Wilson loop2}
\end{align}
In the above expression, we separate the region $\mathcal{M}$ into $\Gamma$ and $\mathcal{M}/\Gamma$ where $\Gamma$ is all the region inside the loop $C$ with the boundary $\partial\Gamma=C$. Denoting the surface area of the region $\Gamma$ and $\mathcal{M}/\Gamma$ by $A_1$ and $A_2$ subsequently together with (\ref{varphi}), the relation (\ref{avg Wilson loop2}) takes the form
\begin{align}
    \langle W[C] \rangle&=\frac{g}{2Z}\sum_\lambda \sum_{N=0}^\infty \exp\bigg[-\bigg(\frac{eg}{2} \bigg)^2 \bigg( A_1(N-\lambda)^2+A_2 N^2 \bigg)\bigg] \label{avg Wilson loop3}.
\end{align}

In the case of SU(2), if we consider the eigenvalues of $\hat{\phi}$ in the fundamental representation, $\lambda=\pm 1/2$. This turns the expression (\ref{avg Wilson loop3}) into
\begin{align}
    \langle W[C] \rangle=&\frac{g}{2Z}\bigg( \sum_{N=-\infty}^\infty \exp\bigg[-e_\text{YM}^2 \bigg( A_1(N+1/2)^2+A_2 N^2 \bigg)\bigg] \nonumber \\&+\exp\bigg[-\frac{e_\text{YM}^2}{4}A_1 \bigg] \bigg) \nonumber \\
    =&\frac{g}{2Z} \bigg( \vartheta\bigg(\frac{ie^2A_1}{2\pi};\frac{ie^2A}{\pi}\bigg)+1\bigg) \exp\bigg[-\frac{e_\text{YM}^2}{4}A_1 \bigg]
    \label{avg Wilson loop4}
\end{align}
where we re-define the Yang-Mills coupling constant $e_\text{YM}$ as $\frac{eg}{2}$ and $\vartheta(z;\tau)$ is the Jacobi's third theta function defined as
\begin{equation}
    \vartheta(z;\tau)=\sum_{N=-\infty}^\infty \exp({2\pi i N z+\pi i N^2 \tau}).
\end{equation}

In the case that $\mathcal{M}$ is an infinitely large sphere, i.e. $A_2\rightarrow\infty$, the vacuum expectation value of the Wilson loop (\ref{avg Wilson loop3}) turns into 
\begin{equation}
    \langle W[C] \rangle=\frac{g}{Z} \exp\bigg[-\frac{e_\text{YM}^2}{4}A_1 \bigg] \label{area law}
\end{equation}
as the theta function becomes unity at this limit.

The result (\ref{area law}) shows that the expectation value of the Wilson loop for 2D Yang-Mills theory obtained by the effective topological BF theory satisfies the area law. This agrees with known results \cite{BRODA1990444,Kondo:1998nw,Kondo:1999tj} as far as the exponent is concerned, which is the dominant piece. To compute the prefactor would require the computing the determinants arising from the Guassian integrals generalising the argument given above for the $SU(2)$ partition function.


\section{Conclusions}
To conclude, we constructed a gauge and Weyl invariant theory of a two-dimensional scalar field by integrating out the gauge fields in BF theory. This model is a candidate for generalising a string theory contact interaction that describes Abelian gauge theory to the non-Abelian case. The calculation was implemented by expanding the fields in the Cartan-Weyl basis. By performing a Gaussian functional integration, we obtained the effective theory with the Lagrangian (\ref{effective}) together with the constraint addressed in (\ref{constraint}). The constraint implies that the magnitude of a scalar field, $|\phi|$, as well as the quantity $d^{ABC}\phi\indices{_A}\phi\indices{_B}\phi\indices{_C}$ are constant throughout the space.

The adjugate and the determinant of the matrix $\widetilde{\mathcal{M}}$ play an important part in (\ref{effective}) where  $\widetilde{\mathcal{M}}$ is defined as (\ref{M}). We developed a diagrammatic approach to represent these objects. The diagrams are constructed from vertices connected to each other by lines. No more than two line are allowed to connect with one vertex. There are two type of diagrams, i.e. a strand and a loop. The adjugate matrix and the matrix determinant were expressed as summations over products of these diagrams as in (\ref{adjugate}) and (\ref{detM}) respectively.

For the case of $SU(2)$ and the manifold having the topology of a unit sphere, the effective action (\ref{SU2 eff2}) contains the winding number of the field $\hat\phi$ which maps a point on the manifold into a point on $S^2$. By using the $SU(2)$ effective action and summing over this winding number, we re-formulated the partition function on a sphere of $SU(2)$ Yang-Mills theory.

Finally, we investigated the BF theory coupled to a source term for the gauge field. The effective generating functional of the gauge field was formulated. The result was checked via a computation of the expectation value of the Wilson loop. We exploited the gauge symmetry to deal with the path-ordering of the Wilson loop. The result showed that the vacuum expectation exhibits the area law agreeing with the well-known results \cite{BRODA1990444,Kondo:1998nw,Kondo:1999tj}.

 \section*{Acknowledgement} We are pleased to acknowledge Development and Promotion of Science and Technology Talents Project (Royal Thai Government Scholarship) for support.


 \bibliographystyle{elsarticle-num} 
 \bibliography{cas-refs}

\begin{thebibliography}{10}
\expandafter\ifx\csname url\endcsname\relax
  \def\url#1{\texttt{#1}}\fi
\expandafter\ifx\csname urlprefix\endcsname\relax\def\urlprefix{URL }\fi
\expandafter\ifx\csname href\endcsname\relax
  \def\href#1#2{#2} \def\path#1{#1}\fi

\bibitem{CHAMSEDDINE1990595}
A.~Chamseddine, D.~Wyler, Topological gravity in 1 + 1 dimensions, Nuclear
  Physics B 340~(2) (1990) 595 -- 616.

\bibitem{Freidel:2012np}
L.~Freidel, S.~Speziale, {On the relations between gravity and BF theories},
  SIGMA 8 (2012) 032.
\newblock \href {http://arxiv.org/abs/1201.4247} {\path{arXiv:1201.4247}},
  \href {https://doi.org/10.3842/SIGMA.2012.032}
  {\path{doi:10.3842/SIGMA.2012.032}}.

\bibitem{Guo:2002yc}
H.-Y. Guo, Y.~Ling, R.-S. Tung, Y.-Z. Zhang, {Chern-Simons term for BF theory
  and gravity as a generalized topological field theory in four-dimensions},
  Phys. Rev. D 66 (2002) 064017.
\newblock \href {http://arxiv.org/abs/hep-th/0204059}
  {\path{arXiv:hep-th/0204059}}, \href
  {https://doi.org/10.1103/PhysRevD.66.064017}
  {\path{doi:10.1103/PhysRevD.66.064017}}.

\bibitem{Smolin:1995vq}
L.~Smolin, {Linking topological quantum field theory and nonperturbative
  quantum gravity}, J. Math. Phys. 36 (1995) 6417--6455.
\newblock \href {http://arxiv.org/abs/gr-qc/9505028}
  {\path{arXiv:gr-qc/9505028}}, \href {https://doi.org/10.1063/1.531251}
  {\path{doi:10.1063/1.531251}}.

\bibitem{WITTEN198846}
E.~Witten, 2 + 1 dimensional gravity as an exactly soluble system, Nuclear
  Physics B 311~(1) (1988) 46 -- 78.

\bibitem{Freidel:1999rr}
L.~Freidel, K.~Krasnov, R.~Puzio, {BF description of higher dimensional gravity
  theories}, Adv. Theor. Math. Phys. 3 (1999) 1289--1324.
\newblock \href {http://arxiv.org/abs/hep-th/9901069}
  {\path{arXiv:hep-th/9901069}}, \href
  {https://doi.org/10.4310/ATMP.1999.v3.n5.a3}
  {\path{doi:10.4310/ATMP.1999.v3.n5.a3}}.

\bibitem{Celada:2016jdt}
M.~Celada, D.~González, M.~Montesinos, {$BF$ gravity}, Class. Quant. Grav.
  33~(21) (2016) 213001.
\newblock \href {http://arxiv.org/abs/1610.02020} {\path{arXiv:1610.02020}},
  \href {https://doi.org/10.1088/0264-9381/33/21/213001}
  {\path{doi:10.1088/0264-9381/33/21/213001}}.

\bibitem{Vishwanath:2012tq}
A.~Vishwanath, T.~Senthil, {Physics of three dimensional bosonic topological
  insulators: Surface Deconfined Criticality and Quantized Magnetoelectric
  Effect}, Phys. Rev. X 3~(1) (2013) 011016.
\newblock \href {http://arxiv.org/abs/1209.3058} {\path{arXiv:1209.3058}},
  \href {https://doi.org/10.1103/PhysRevX.3.011016}
  {\path{doi:10.1103/PhysRevX.3.011016}}.

\bibitem{Marzuoli_2012}
A.~Marzuoli, G.~Palumbo, {BF}-theory in graphene: A route toward topological
  quantum computing?, {EPL} (Europhysics Letters) 99~(1) (2012) 10002.
\newblock \href {https://doi.org/10.1209/0295-5075/99/10002}
  {\path{doi:10.1209/0295-5075/99/10002}}.

\bibitem{Palumbo:2013gxa}
G.~Palumbo, R.~Catenacci, A.~Marzuoli, {Topological effective field theories
  for Dirac fermions from index theorem}, Int. J. Mod. Phys. B 28 (2014)
  1350193.
\newblock \href {http://arxiv.org/abs/1303.6468} {\path{arXiv:1303.6468}},
  \href {https://doi.org/10.1142/S0217979213501932}
  {\path{doi:10.1142/S0217979213501932}}.

\bibitem{Cho:2010rk}
G.~Y. Cho, J.~E. Moore, {Topological BF field theory description of topological
  insulators}, Annals Phys. 326 (2011) 1515--1535.
\newblock \href {http://arxiv.org/abs/1011.3485} {\path{arXiv:1011.3485}},
  \href {https://doi.org/10.1016/j.aop.2010.12.011}
  {\path{doi:10.1016/j.aop.2010.12.011}}.

\bibitem{Thakurathi:2020veg}
M.~Thakurathi, A.~Burkov, {Theory of the fractional quantum Hall effect in Weyl
  semimetals} (5 2020).
\newblock \href {http://arxiv.org/abs/2005.13545} {\path{arXiv:2005.13545}}.

\bibitem{Blasi:2011pf}
A.~Blasi, A.~Braggio, M.~Carrega, D.~Ferraro, N.~Maggiore, N.~Magnoli,
  {Non-Abelian BF theory for 2+1 dimensional topological states of matter}, New
  J. Phys. 14 (2012) 013060.
\newblock \href {http://arxiv.org/abs/1106.4641} {\path{arXiv:1106.4641}},
  \href {https://doi.org/10.1088/1367-2630/14/1/013060}
  {\path{doi:10.1088/1367-2630/14/1/013060}}.

\bibitem{You2019FractonicCA}
Y.~You, T.~Devakul, S.~L. Sondhi, F.~J. Burnell, Fractonic chern-simons and bf
  theories, arXiv: Strongly Correlated Electrons (2019).

\bibitem{Mansfield:2011eq}
P.~Mansfield, {Faraday's Lines of Force as Strings: from Gauss' Law to the
  Arrow of Time}, JHEP 10 (2012) 149.
\newblock \href {http://arxiv.org/abs/1108.5094} {\path{arXiv:1108.5094}},
  \href {https://doi.org/10.1007/JHEP10(2012)149}
  {\path{doi:10.1007/JHEP10(2012)149}}.

\bibitem{Edwards:2014cga}
J.~P. Edwards, P.~Mansfield, {QED as the tensionless limit of the spinning
  string with contact interaction}, Phys. Lett. B 746 (2015) 335--340.
\newblock \href {http://arxiv.org/abs/1409.4948} {\path{arXiv:1409.4948}},
  \href {https://doi.org/10.1016/j.physletb.2015.05.024}
  {\path{doi:10.1016/j.physletb.2015.05.024}}.

\bibitem{Edwards:2014xfa}
J.~P. Edwards, P.~Mansfield, {Delta-function Interactions for the Bosonic and
  Spinning Strings and the Generation of Abelian Gauge Theory}, JHEP 01 (2015)
  127.
\newblock \href {http://arxiv.org/abs/1410.3288} {\path{arXiv:1410.3288}},
  \href {https://doi.org/10.1007/JHEP01(2015)127}
  {\path{doi:10.1007/JHEP01(2015)127}}.

\bibitem{witten1991}
E.~Witten, On quantum gauge theories in two dimensions, Comm. Math. Phys.
  141~(1) (1991) 153--209.

\bibitem{Witten:1992xu}
E.~Witten, {Two-dimensional gauge theories revisited}, J. Geom. Phys. 9 (1992)
  303--368.
\newblock \href {http://arxiv.org/abs/hep-th/9204083}
  {\path{arXiv:hep-th/9204083}}, \href
  {https://doi.org/10.1016/0393-0440(92)90034-X}
  {\path{doi:10.1016/0393-0440(92)90034-X}}.

\bibitem{Rusakov:1990rs}
B.~Rusakov, {Loop averages and partition functions in U(N) gauge theory on
  two-dimensional manifolds}, Mod. Phys. Lett. A 5 (1990) 693--703.
\newblock \href {https://doi.org/10.1142/S0217732390000780}
  {\path{doi:10.1142/S0217732390000780}}.

\bibitem{Samuel:1978iy}
S.~Samuel, {COLOR ZITTERBEWEGUNG}, Nucl. Phys. B 149 (1979) 517--524.
\newblock \href {https://doi.org/10.1016/0550-3213(79)90005-1}
  {\path{doi:10.1016/0550-3213(79)90005-1}}.

\bibitem{Broda:1995wv}
B.~Broda, {NonAbelian Stokes theorem} (1995) 496--505\href
  {http://arxiv.org/abs/hep-th/9511150} {\path{arXiv:hep-th/9511150}}.

\bibitem{Curry:2017cnu}
C.~Curry, P.~Mansfield, {Intersection of world-lines on curved surfaces and
  path-ordering of the Wilson loop}, JHEP 06 (2018) 081.
\newblock \href {http://arxiv.org/abs/1712.04760} {\path{arXiv:1712.04760}},
  \href {https://doi.org/10.1007/JHEP06(2018)081}
  {\path{doi:10.1007/JHEP06(2018)081}}.

\bibitem{BRODA1990444}
B.~Broda, Two-dimensional topological yang-mills theory, Physics Letters B
  244~(3) (1990) 444 -- 449.

\bibitem{Kondo:1998nw}
K.-I. Kondo, {Abelian magnetic monopole dominance in quark confinement}, Phys.
  Rev. D 58 (1998) 105016.
\newblock \href {http://arxiv.org/abs/hep-th/9805153}
  {\path{arXiv:hep-th/9805153}}, \href
  {https://doi.org/10.1103/PhysRevD.58.105016}
  {\path{doi:10.1103/PhysRevD.58.105016}}.

\bibitem{Kondo:1999tj}
K.-I. Kondo, Y.~Taira, {NonAbelian Stokes theorem and quark confinement in
  SU(N) Yang-Mills gauge theory}, Prog. Theor. Phys. 104 (2000) 1189--1265.
\newblock \href {http://arxiv.org/abs/hep-th/9911242}
  {\path{arXiv:hep-th/9911242}}, \href {https://doi.org/10.1143/PTP.104.1189}
  {\path{doi:10.1143/PTP.104.1189}}.

\end{thebibliography}






\newpage
\begin{landscape}
\begin{appendices}
\section{Expressions for Matrix Multiplications of the Matrix $\mathcal{F}$}
The expressions for matrix multiplication of the matrix $\mathcal{F}$ are given as follows:
\begin{align}
    &\mathcal{F}\indices{^A_B}\mathcal{F}\indices{^B^C}= \nonumber \\
    &\begin{pmatrix}
    \phantom{00}0\phantom{00}&  \phantom{00}0\phantom{00}&  0&  0&  0&  0&  0&  0 \\
    0&  0&  0&  0&  0&  0&  0&  0 \\
    0&  0&  0&  -2(\phi\indices{_1})^2&  0&  0&  0&  0 \\
    0&  0&  -2(\phi\indices{_1})^2&  0&   0&  0&  0&  0 \\
    0&  0&  0&  0&  0&  \frac{-1}{2}(\phi\indices{_1}-2\phi\indices{_2})^2&  0&  0 \\
    0&  0&  0&  0&  \frac{-1}{2}(\phi\indices{_1}-2\phi\indices{_2})^2& 0 &  0&  0 \\
    0&  0&  0&  0&  0&  0&  0&  \frac{-1}{2}(\phi\indices{_1}+2\phi\indices{_2})^2 \\
    0&  0&  0&  0&  0&  0&  \frac{-1}{2}(\phi\indices{_1}+2\phi\indices{_2})^2&  0 \\
    \end{pmatrix} 
\end{align}

\begin{align}
    &\mathcal{F}\indices{^A_B}\mathcal{F}\indices{^B_C}\mathcal{F}^{CD}= \nonumber \\ 
    &\begin{pmatrix}
    \phantom{00}0\phantom{00}&  \phantom{00}0\phantom{00}&  0&  0&  0&  0&  0&  0 \\
    0&  0&  0&  0&  0&  0&  0&  0 \\
    0&  0&  0&  -2i(\phi\indices{_1})^3&  0&  0&  0&  0 \\
    0&  0&  2i(\phi\indices{_1})^3&  0&   0&  0&  0&  0 \\
    0&  0&  0&  0&  0&  \frac{i}{4}(\phi\indices{_1}-2\phi\indices{_2})^3&  0&  0 \\
    0&  0&  0&  0&  \frac{-i}{4}(\phi\indices{_1}-2\phi\indices{_2})^3& 0 &  0&  0 \\
    0&  0&  0&  0&  0&  0&  0&  \frac{-i}{4}(\phi\indices{_1}+2\phi\indices{_2})^3 \\
    0&  0&  0&  0&  0&  0&  \frac{i}{4}(\phi\indices{_1}+2\phi\indices{_2})^3&  0 \\
    \end{pmatrix} 
\end{align}

\begin{align}
    &\mathcal{F}\indices{^A_B}\mathcal{F}\indices{^B_C}\mathcal{F}\indices{^C_D}\mathcal{F}\indices{^D^E}= \nonumber \\
    &\begin{pmatrix}
    \phantom{00}0\phantom{00}&  \phantom{00}0\phantom{00}&  0&  0&  0&  0&  0&  0 \\
    0&  0&  0&  0&  0&  0&  0&  0 \\
    0&  0&  0&  2(\phi\indices{_1})^4&  0&  0&  0&  0 \\
    0&  0&  2(\phi\indices{_1})^4&  0&   0&  0&  0&  0 \\
    0&  0&  0&  0&  0&  \frac{1}{8}(\phi\indices{_1}-2\phi\indices{_2})^4&  0&  0 \\
    0&  0&  0&  0&  \frac{1}{8}(\phi\indices{_1}-2\phi\indices{_2})^4& 0 &  0&  0 \\
    0&  0&  0&  0&  0&  0&  0&  \frac{1}{8}(\phi\indices{_1}+2\phi\indices{_2})^4 \\
    0&  0&  0&  0&  0&  0&  \frac{1}{8}(\phi\indices{_1}+2\phi\indices{_2})^4&  0 \\
    \end{pmatrix} 
\end{align}

\begin{align}
    &\mathcal{F}\indices{^A_B}\mathcal{F}\indices{^B_C}\mathcal{F}\indices{^C_D}\mathcal{F}\indices{^D_E}\mathcal{F}\indices{^E^F}= \nonumber \\
    &\begin{pmatrix}
    \phantom{00}0\phantom{00}&  \phantom{00}0\phantom{00}&  0&  0&  0&  0&  0&  0 \\
    0&  0&  0&  0&  0&  0&  0&  0 \\
    0&  0&  0&  2i(\phi\indices{_1})^5&  0&  0&  0&  0 \\
    0&  0&  -2i(\phi\indices{_1})^5&  0&   0&  0&  0&  0 \\
    0&  0&  0&  0&  0&  \frac{-i}{16}(\phi\indices{_1}-2\phi\indices{_2})^5&  0&  0 \\
    0&  0&  0&  0&  \frac{i}{16}(\phi\indices{_1}-2\phi\indices{_2})^5& 0 &  0&  0 \\
    0&  0&  0&  0&  0&  0&  0&  \frac{i}{16}(\phi\indices{_1}+2\phi\indices{_2})^5 \\
    0&  0&  0&  0&  0&  0&  \frac{-i}{16}(\phi\indices{_1}+2\phi\indices{_2})^5&  0 \\
    \end{pmatrix} 
\end{align}

\begin{align}
    &\mathcal{F}\indices{^A_B}\mathcal{F}\indices{^B_C}\mathcal{F}\indices{^C_D}\mathcal{F}\indices{^D_E}\mathcal{F}\indices{^E_F}\mathcal{F}\indices{^F^G}= \nonumber \\
    &\begin{pmatrix}
    \phantom{00}0\phantom{00}&  \phantom{00}0\phantom{00}&  0&  0&  0&  0&  0&  0 \\
    0&  0&  0&  0&  0&  0&  0&  0 \\
    0&  0&  0&  -2(\phi\indices{_1})^6&  0&  0&  0&  0 \\
    0&  0&  -2(\phi\indices{_1})^6&  0&   0&  0&  0&  0 \\
    0&  0&  0&  0&  0&  \frac{-1}{32}(\phi\indices{_1}-2\phi\indices{_2})^6&  0&  0 \\
    0&  0&  0&  0&  \frac{-1}{32}(\phi\indices{_1}-2\phi\indices{_2})^6& 0 &  0&  0 \\
    0&  0&  0&  0&  0&  0&  0&  \frac{-1}{32}(\phi\indices{_1}+2\phi\indices{_2})^6 \\
    0&  0&  0&  0&  0&  0&  \frac{-1}{32}(\phi\indices{_1}+2\phi\indices{_2})^6&  0 \\
    \end{pmatrix} 
\end{align}

\end{appendices}
\end{landscape}

\end{document}